\documentclass[journal]{IEEEtran}
\pdfoutput=1
\renewcommand{\vec}[1]{\ensuremath{\boldsymbol{#1}}} 
\usepackage[german,english]{babel}
\usepackage{amssymb}
\usepackage[latin1]{inputenc}
\usepackage{graphicx}
\usepackage{eurosym}
\usepackage{pstool}
\usepackage{acronym}
\usepackage[usenames,dvipsnames]{pstricks}
\usepackage{epsfig}
\usepackage{pst-grad} 
\usepackage{pst-plot} 
\usepackage{multirow}
\usepackage{threeparttable}
\usepackage{pstool}
\usepackage{relsize}
\usepackage{siunitx}
\usepackage[caption=false,font=footnotesize]{subfig}
\DeclareSIUnit \voltampere { VA } 
\DeclareSIUnit \watthour { Wh } 

\usepackage{algorithm}
\usepackage{algpseudocode}
\algdef{SE}[DOWHILE]{Do}{doWhile}{\algorithmicdo}[1]{\algorithmicwhile\ #1}%

\usepackage[cmex10]{amsmath}
\usepackage{tabularx}
\newcommand{\PreserveBackslash}[1]{\let\temp=\\#1\let\\=\temp}

\begingroup
\makeatletter
\catcode`\#=11
    \gdef\pstool@bitmap@opts{%
      -dAutoFilterColorImages#false
      -dAutoFilterGrayImages#false %
      -dColorImageFilter#/FlateEncode %
      -dGrayImageFilter#/FlateEncode 
      }
    \gdef\pstool@pspdf@opts{%
      -dPDFSETTINGS#/prepress %
      -dCompatibilityLevel#1.3 %
      -dEmbedAllFonts#true%
      -dSubsetFonts#true
      }
\endgroup

\usepackage{mathdots}

\ifCLASSINFOpdf
\else
\fi
\usepackage{array}
\usepackage{fixltx2e}

\usepackage{cite}

\begin{document}
%
\title{Modeling and Optimal Operation of Distributed Battery Storage in Low Voltage Grids}



\author{ \IEEEauthorblockA{Philipp Fortenbacher, Johanna L. Mathieu, and  G\"oran Andersson
	}
	\\ 
	\thanks{This work was, in part, financed by the Swiss Commission for Technology and Innovation (CTI) 4/2013-10/2014
		
		P. Fortenbacher and G. Andersson  are with the Power Systems Laboratory, ETH Zurich, Switzerland (e-mail: fortenbacher@eeh.ee.ethz.ch, andersson@eeh.ee.ethz.ch). 
		
		J. L. Mathieu is with the Department of Electrical Engineering and Computer Science, University of Michigan, USA (e-mail: jlmath@umich.edu).}
}


\acrodef{LV}[LV]{Low Voltage}
\acrodef{AC-OPF}[AC-OPF]{AC Optimal Power Flow}
\acrodef{OPF}[OPF]{Optimal Power Flow}
\acrodef{FBS-OPF}[FBS-OPF]{Forward Backward Sweep Optimal Power Flow}
\acrodef{FBS}[FBS]{Forward Backward Sweep}
\acrodef{IP}[IP]{Interior Point}
\acrodef{LP}[LP]{Linear Programming}
\acrodef{SOCP}[SOCP]{Second Order Cone Programming}
\acrodef{SDP}[SDP]{Semi Definite Programming}
\acrodef{etaload}[$\eta_{\mathrm{load}}$] {charge efficiency}
\acrodef{etagen}[$\eta_{\mathrm{gen}}$] {discharge efficiency}
\acrodef{etabat}[$\eta_{\mathrm{bat}}$] {battery efficiency}
\acrodef{etain}[$\eta_{\mathrm{in}}$] {converter efficiency}
\acrodef{etatot}[$\eta_{\mathrm{tot}}$] {total battery system efficiency}
\acrodef{ploss}[$P^{\mathrm{loss}}_\mathrm{bat}$]{battery loss power}
\acrodef{pmax}[$P_{\mathrm{max}}$]{max battery power}
\acrodef{R}[$R$] {internal resistance}
\acrodef{cw}[$c_{\mathrm{w}}$] {charge well factor}
\acrodef{SOC}[SOC]{state of charge}
\acrodef{cr}[$c_{\mathrm{r}}$] {recovery factor}
\acrodef{Q}[$Q_{\mathrm{bat}}$] {charge capacity}
\acrodef{C}[$C_{\mathrm{bat}}$] {energy capacity}
\acrodef{Vocavg}[$\bar{V}_{\mathrm{oc}}$] {average open circuit potential}
\acrodef{Vocsoc}[$V_{\mathrm{oc}}$] {average open circuit potential}
\acrodef{Is}[$I_{\mathrm{s}}$]{side current}
\acrodef{vt}[$V_{\mathrm{t}}$]{terminal voltage}
\acrodef{ibat}[$I_{\mathrm{bat}}$]{battery current}
\acrodef{pbat}[$P_{\mathrm{bat}}$]{battery power}
\acrodef{cost}[$c_\mathrm{inv}$]{investment cost}
\acrodef{x}[$x$]{state variable for non-linear and linear battery models. Corresponds to SOC}
\acrodef{x_ex}[$\vec{x}$]{state vector containing $[x_1,x_2]^T$ for the extended non-linear and linear models.}
\acrodef{a}[$a$]{SOC target}
\acrodef{b}[$b$]{cost parameter SOC}
\acrodef{c}[$c$]{cost parameter $u_{\mathrm{gen}}$}
\acrodef{d}[$d$]{cost parameter $u_{\mathrm{load}}$}
\acrodef{e}[$e$]{cost parameter $u_{\mathrm{load}}^2$}
\acrodef{cnet}[$c_{\mathrm{net}}$]{net tariff}
\acrodef{cinv}[$c_{\mathrm{inv}}$]{investment cost}
\acrodef{ncell}[$n_{\mathrm{cell}}$]{number of DUALFOIL cells in series}
\acrodef{Acell}[$A_{\mathrm{cell}}$]{cell area}
\acrodef{dclink}[$v_{\mathrm{DC}}$]{DC link voltage}
\acrodef{pthreshold}[$u_{\mathrm{gen,net}}^{\mathrm{max}}$]{peak shave threshold}
\acrodef{horizonT}[$H_{\mathrm{t}}$]{time horizon}
\acrodef{updateT}[$R_{\mathrm{t}}$]{receding horizon window}
\acrodef{sampleTime}[$T_{\mathrm{s}}$]{sample rate}
\acrodef{rsa}[$r_{\mathrm{sa}}$]{side reaction rate constant anode}
\acrodef{rsc}[$r_{\mathrm{sc}}$]{side reaction rate constant cathode}
\acrodef{unet}[$u_{\mathrm{gen}}^{\mathrm{net}}$]{}
\acrodef{unetmax}[$u_{\mathrm{gen,max}}^{\mathrm{net}}$]{}
\acrodef{ubatgen}[$u_{\mathrm{gen}}^{\mathrm{bat}}$]{}
\acrodef{ubatload}[$u_{\mathrm{load}}^{\mathrm{bat}}$]{}
\acrodef{uloadG2}[$P_{\mathrm{load}}^{\mathrm{G2}}$]{G2 standard industrial load profile}
\acrodef{Qs}[$Q_{\mathrm{s}}$]{lost charge}

\acrodef{ID}[ID]{Identification}
\acrodef{NLS}[NLS]{Nonlinear Least Squares}
\acrodef{LS}[LS]{Least Squares}
\acrodef{MPC}[MPC]{Model Predictive Control}
\acrodef{QP}[QP]{Quadratic Programming}
\acrodef{MSE}[MSE]{Mean Squared Error}
\acrodef{RMSE}[RMSE]{Root Mean Squared Error}
\acrodef{NRMSE}[NRMSE]{Normalized Root Mean Squared Error}
\acrodef{PWA}[PWA]{Piece-Wise Affine}
\acrodef{DOD}[DOD]{Depth Of Discharge}
\acrodef{PFC}[PFC]{Primary Frequency Control}
\acrodef{LFC}[LFC]{Load Frequency Control}
\acrodef{MIQP}[MIQP]{Mixed Integer Quadratic Programming}
\acrodef{MINLP}[MINLP]{Mixed Integer Non Linear Programming}
\acrodef{FC}[FC]{Frequency Control}
\acrodef{SOS}[SOS]{Special Ordered Set}
\acrodef{PV}[PV]{photovaltaics}
\acrodef{ARX}[ARX]{AutoRegressive with eXogenous input}
\acrodef{DSO}[DSO]{Distribution System Operator}
\acrodef{PWA}[PWA]{piecewise-affine}
\acrodef{MILP}[MILP]{Mixed Integer Linear Programming}
\acrodef{SOS}[SOS]{Special Ordered Set}
\acrodef{RT}[RT]{real-time}
\acrodef{SoE}[SoE]{State of Energy}

\maketitle

\begin{abstract}
Due to high power in-feed from photovoltaics, it can be expected that more battery systems will be installed in the distribution grid in near future to mitigate voltage violations and thermal line and transformer overloading. In this paper, we present a two-stage centralized model predictive control scheme for distributed battery storage that consists of a scheduling entity and a real-time control entity. To guarantee secure grid operation, we solve a robust multi-period optimal power flow (OPF) for the scheduling stage that minimizes battery degradation and maximizes photovoltaic utilization subject to grid constraints. The real-time controller solves a real-time OPF taking into account storage allocation profiles from the scheduler, a detailed battery model, and real-time measurements. To reduce the computational complexity of the controllers, we present a linearized OPF that approximates the nonlinear AC-OPF into a linear programming problem. Through a case study, we show, for two different battery technologies, that we can substantially reduce battery degradation when we incorporate a battery degradation model. A further finding is that we can reduce battery losses by 30\% by using the detailed battery model in the real-time control stage.
\end{abstract}


\section*{Nomenclature}
\addcontentsline{toc}{section}{Nomenclature}
\begin{IEEEdescription}[\IEEEusemathlabelsep\IEEEsetlabelwidth{$\vec{A}_1,\vec{A}_2,\vec{A}_3,t$}]

	\item[$\eta_\mathrm{bat},\eta_\mathrm{in}$] battery stack and inverter efficiency
	\item[$\eta_\mathrm{bat}^\mathrm{dis},\eta_\mathrm{bat}^\mathrm{ch}$] battery stack discharging and charging efficiency
	\item[$\eta_\mathrm{dis},\eta_\mathrm{ch}$] total battery system discharging and charging efficiency  
	\item[$\vec{\lambda}$] SOS2 set
	\item[$\vec{\sigma}_\mathrm{PV}$] standard deviation of the PV forecast error
	\item[$\vec{\Phi}$] discrete battery system dynamics matrix
	
	\item[$\vec{a}_1,\vec{a}_2,\vec{a}_3$] degradation plane parameter vectors of the triangles from the convex hull
	\item[$\vec{A}$] continuous battery system dynamics matrix
	\item[$\vec{A}_1,\vec{A}_2,\vec{A}_3,\vec{A}_\mathrm{z}$] matrices to include battery degradation for multiple battery systems and time steps
	\item[$\vec{A}_q$] matrix to describe polygonal regions for active and reactive power

	\item[$\vec{b}_\mathrm{bat}$] battery system control input vector
	\item[$\vec{B}$] battery system control input matrix for multiple battery systems
	\item[$\vec{B}_\mathrm{i}^\mathrm{1}, \vec{B}_\mathrm{i}^\mathrm{2}, \vec{b}_\mathrm{l}$] Matrices and vector specifying the PWA network loss approximation 
	\item[$\vec{B}_\mathrm{r}$] branch flow matrix
	\item[$\vec{B}_\mathrm{v}$] linearized active and reactive power to voltage matrix
	\item[$\vec{B}_q$] matrix to describe polygonal regions for active and reactive power
	
	\item[$\vec{C}_\mathrm{g}$] controllable generator to bus mapping matrix
	\item[$\vec{C}_\mathrm{pv}$] PV generator to bus mapping matrix
	
	\item[$c_\mathrm{E}$] battery capacity
	\item[$\vec{c}_\mathrm{E}$] battery capacity vector
	\item[$c_\mathrm{d}$] battery degradation cost in \euro/kWh
	\item[$c_\mathrm{n}$] feeder energy cost in \euro/MWh
	\item[$\vec{c}_\mathrm{p}$] generator energy cost vector
	\item[$c_\mathrm{r}$] inverse charge recovery time in sec$^{-1}$
	\item[$c_\mathrm{w}$] parameter to describe the size of capacity wells
	
	\item[$E$] state of energy of one battery system in MWh
	\item[$\vec{e}$] state of energy vector of $n_s$ battery systems
	\item[$\vec{e}(0)$] initial state of energy vector of $n_s$ battery systems
	\item[$\vec{e}_\mathrm{min},\vec{e}_\mathrm{max}$] storage allocation bounds from scheduler	
	\item[$\vec{E}$] state of energy evolution vector of $n_s$ battery systems
	
	\item[$\vec{H}$] discrete battery control input matrix

	\item[$\vec{i}$] nodal current vector in p.u.
	\item[$\vec{i}_\mathrm{b}$] branch current vector in p.u.
	\item[$\vec{i}_\mathrm{b}^{\mathrm{max}}$] max branch current vector in p.u.
	\item[$I_\mathrm{bat}$] battery current in A
	
	\item[$\vec{l}$] Luenberger observer gain
	
	\item[$m$] penalty factor for setpoint regions 
	\item[$n$] number of buses
	\item[$n_\mathrm{g}$] number of controllable generators
	\item[$n_\mathrm{l}$] number of branches
	\item[$n_\mathrm{p}$] number of triangles
	\item[$n_\mathrm{pv}$] number of PV units
	\item[$n_s$] number of battery systems
	\item[$N$] number of time steps in the scheduler horizon
	
	\item[$P_\mathrm{bat}$] active battery stack power
	\item[$p_\mathrm{bat}^{\mathrm{agg}}$] aggregated battery set point
	\item[${P}_\mathrm{bat}^{\mathrm{loss}}$] total active power losses of one battery system
	\item[$\vec{p}_\mathrm{bat}^{\mathrm{loss}}$] total active power loss vector of $n_s$ battery systems
	\item[${P}_\mathrm{bat,r}^+, {P}_\mathrm{bat,r}^-$] active battery power range for linear power loss approximation 
	\item[$P_\mathrm{cell}$] active battery cell power
	\item[$\vec{p}_\mathrm{d},\vec{q}_\mathrm{d}$] non-controllable nodal active and reactive power load vectors
	\item[$\vec{p}_\mathrm{gen},\vec{q}_\mathrm{gen}$] controllable active and reactive generator power vectors
	\item[$\vec{p}_\mathrm{gen}^{\mathrm{pv}},\vec{q}_\mathrm{gen}^{\mathrm{pv}},\hat{\vec{p}}_\mathrm{gen}^{\mathrm{pv}}$] active and reactive power PV measurement and prediction vectors
	\item[$\vec{p}_\mathrm{gen}^{\mathrm{s,dis}} \in \vec{p}_\mathrm{gen}^{\mathrm{s}}$] active discharging battery system grid power vector
	\item[$\vec{p}_\mathrm{gen}^{\mathrm{s,ch}} \in \vec{p}_\mathrm{gen}^{\mathrm{s}}$] active charging battery system grid power vector
	\item[$\vec{p}_\mathrm{gen}^{\mathrm{s},P}$] supporting point vector for nonconvex battery system loss calculation
	\item[$\vec{p}_\mathrm{l}^{\mathrm{p}},\vec{p}_\mathrm{l}^{\mathrm{q}}$] decision vectors of active network losses
	\item[$\vec{p}_\mathrm{ld},\vec{q}_\mathrm{ld},\hat{\vec{p}}_\mathrm{ld}$] active and reactive power load measurement and prediction vectors
	\item[$\vec{p}_\mathrm{min}, \vec{p}_\mathrm{max}$] min and max active generator power vectors
	\item[$p_\mathrm{net}$] active feeder generator power
		
	\item[$R$] internal battery resistance
	\item[$\vec{s}_\mathrm{max}$] max apparent generator power vector
	\item[$\vec{S}_x,\vec{S}_u$] matrices to describe the storage evolution

	\item[$T_1,T_2$] sample times for scheduler and RT controller
	\item[$\vec{U}$] decision vector for active battery power
	\item[$\vec{v}$] nodal RMS voltage vector in p.u.
	\item[$\vec{v}_\mathrm{min}, \vec{v}_\mathrm{max}$] min and max nodal RMS voltage vectors
	\item[$\vec{v}_\mathrm{s}$] slack bus voltage vector
	\item[$V_\mathrm{oc}$] battery stack open-circuit voltage
	\item[$\vec{w}_k \in \vec{W}$]  box-constrained uncertainty set
	\item[$\vec{x}_k \in \vec{X}$] decision vector for grid variables
	\item[$\vec{x}_\mathrm{E}$] dynamic state vector of one battery system
	\item[${x}_\mathrm{E1}$]  state of accessible energy well
	\item[${x}_\mathrm{E2}$]  state of non-accessible energy well
	\item[$\vec{x}_\mathrm{set}$] vector of helper decision variables to penalize deviations from the storage allocation band
	\item[$\vec{X}_0,\vec{X}_1$] battery system state vectors (initial and after one time step) for $n_s$ battery systems
	\item[$y_\mathrm{set}^{\mathrm{agg}}$]  helper decision variable to penalize deviations from the aggregated battery set point
	\item[$z$] battery degradation variable for one battery system 
	\item[$\vec{z}_k \in \vec{Z}$] decision vector for battery degradation variables
\end{IEEEdescription}


\section{Introduction}
\IEEEPARstart{I}{n} the next few years, it can be expected that many battery systems will be installed in the \ac{LV} distribution grid to cope with high in-feed from \ac{PV} \cite{eurobat} and other fluctuating energy sources also connected to the distribution grid. In particular, battery systems can mitigate voltage violations and thermal line overloading, allowing \acp{DSO} to defer line and transformer upgrades. Some recent papers \cite{Marra2014,Chua2012} developed decentralized battery control strategies to provide voltage support. Decentralized control strategies have the advantage that they rely only on local measurements. Thus, they do not require communication infrastructure. However, addressing thermal overloading requires coordination either via centralized or distributed control, since the power flows are not observable at a local level.  In \cite{Wang2014}, a centralized predictive control scheme is developed to avoid thermal line overloading, but the authors do not consider voltage constraints. Furthermore, the proposed control strategies in \cite{Marra2014,Chua2012,Wang2014} do not consider battery degradation or use detailed battery models.

Since investment costs for batteries are still high \cite{Nykvist2015}, a battery's expected lifetime greatly affects assessments of its economic viability. Each control action applied to a battery leads to charge capacity loss (i.e., degradation) \cite{Moura2011}, reducing its lifetime. Some recent papers propose methods to include battery degradation costs in economic cost functions \cite{koller,trippe}. Another way to increase the economic viability of a battery is to exploit its full capacity, enabling operation in low/high state of charge regimes when the benefits outweigh the degradation costs. However, simple battery models do not capture the dynamics that are present when batteries operate in these regimes. For example, operation in these regimes is only possible at low charging/ discharging powers. It is possible to model these dynamics with detailed battery models \cite{fortenbacherPSCC14,fortenbacherPowerTech2015}.

The objective of this paper is to develop computationally-tractable methods to control distributed batteries in distribution networks with high penetrations of PV to manage network constraints such as voltage and thermal constraints. To exploit the full potential of the network and to dispatch the battery systems in an economic and efficient way, we formulate multi-period \ac{AC-OPF} problems. We use a linear approximation of the nonlinear nonconvex \ac{AC-OPF} from our previous work \cite{fortenbacherPSCC16} that exploits the radial structure of a \ac{LV} network. The approximation captures line losses and is linear in active and reactive power injections. Linear approximations have been developed for distribution networks \cite{Gan2014,BaranWu1989}; however, these neglect line losses. The so-called DC power flow approximation and the approximation in \cite{Taylor2011} are not applicable for distribution grids since, in LV grids, active power flow is dominated by voltage magnitude differences rather than voltage angle differences. Nonlinear convex relaxations of the \ac{AC-OPF} problem also exist; however, the resulting semi definite \cite{molzahn,DallAnese2014} and second order cone \cite{low_socp,Christakou2015} programs are computationally large and not suitable for computationally-limited controllers. Linear programming approximations of the second order cone programming problem have been developed, but the number of the linear constraints is large \cite{Mhanna2016}. Ref.~\cite{Bolognani2016} derives an AC power flow approximation for distribution networks that is linear between the complex voltage and the complex power injections, but does not demonstrate how to use the approximation within an \ac{OPF} problem.

Our contribution is the development of a control strategy that leverages the linearized \ac{AC-OPF} approximation to optimize distributed battery operation within a LV grid. We incorporate the linear \ac{AC-OPF} into a two-stage \ac{MPC} control scheme that consists of a scheduler and a \ac{RT} controller. The scheduler is a robust \ac{MPC} that solves a multi-period \ac{OPF} minimizing battery degradation and maximizing PV utilization subject to grid and storage constraints. We use the concept of degradation maps allowing us to describe the impact on degradation associated with discrete control actions. By using a \ac{PWA} representation of the map, we are able to use efficient \ac{LP} solvers. We link the planning domain with the \ac{RT} domain by computing storage allocation bounds that are used by the \ac{RT} controller, which solves an \ac{RT}-\ac{OPF} using a detailed battery model and \ac{RT} measurements of the network and battery states to minimize the battery and network losses. As a further contribution, we show that simple battery models lead to high power errors within \ac{RT} control applications.

This paper is organized as follows: Section~\ref{sec:problemDef} defines the problem that we aim to solve. Section~\ref{sec:batmodeling} presents the battery models used in the proposed controllers. Section~\ref{sec:opf} reviews the linearized \ac{OPF} formulation that is included in our controllers. Section~\ref{sec:control} describes the controller formulations for the scheduler and \ac{RT} controller. Section~\ref{sec:results} presents the simulation results and a battery lifetime assessment. Finally, Section~\ref{sec:conclusion} presents the conclusions.

\section{Problem Definition}
\label{sec:problemDef}

Existing inverter-level strategies to maximize PV infeed include active power curtailment \cite{tonkoski2011coordinated} and reactive power control \cite{kerber}, which are used by \acp{DSO} in Germany \cite{VDE-AR-N}. In contrast to \ac{OPF} methods, no communication infrastructure is needed. However, OPF methods enable optimal utilization of the grid. We propose a method that combines distribution-level \ac{OPF} with a storage control strategy that could leverage  high bandwidth communication links already available at most households.

\begin{figure}
	\centering
	\def\svgwidth{\columnwidth}
	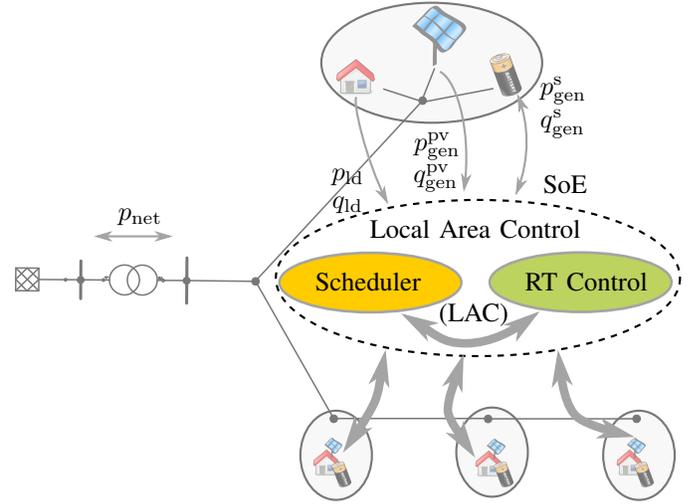
	\caption{Illustration of the problem environment. The overall control objective is to maximize PV infeed while managing grid constraints and minimizing battery degradation.}
	\label{fig:problemDef}
\end{figure}

Figure~\ref{fig:problemDef} illustrates the problem environment. Battery systems are used to mitigate voltage violations and thermal line overloading that results from high PV infeed in \ac{LV} networks, enabling the DSO to defer equipment upgrades. The overall control objective is to maximize PV infeed, while managing grid constraints and minimizing battery degradation. \ac{PV} maximization can be achieved by minimizing the net power $p_\mathrm{net}$. We assume local area control (LAC) entities compute optimal allocation schedules for the storage devices in their area every hour for a 24 hour horizon. We assume each area contains a fraction of the total number of storage devices within the full distribution network, for example, 10-200 devices connected to a feeder. It uses the storage allocation schedule, knowledge of each battery's \ac{SoE}, and \ac{RT} measurements of the active and reactive power consumption of the loads ($p_\mathrm{ld}, q_\mathrm{ld}$) and power production of the \ac{PV} generators ($p_\mathrm{gen}^{\mathrm{pv}}, q_\mathrm{gen}^{\mathrm{pv}}$) to solve an \ac{RT}-\ac{OPF} to compute \ac{RT} active and reactive power battery setpoints ($p_\mathrm{gen}^{\mathrm{s}}, q_\mathrm{gen}^{\mathrm{s}}$) every 10 seconds.  

\begin{figure}
	\centering
	\def\svgwidth{\columnwidth}
	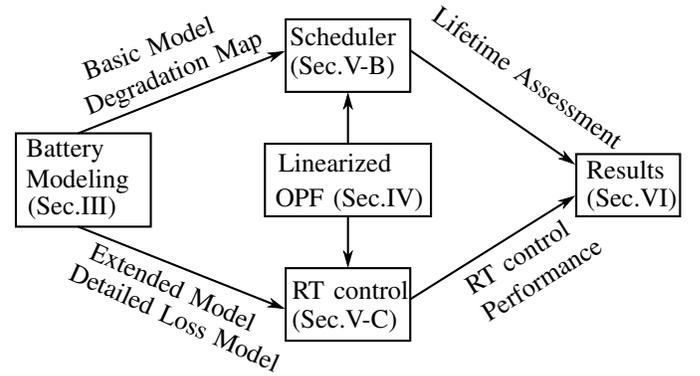
	\caption{Relationships between the sections of the paper.}
	\label{fig:overview}
\end{figure}

Figure~\ref{fig:overview} shows the relationships between the sections of the paper. Different battery models are presented in Section~\ref{sec:batmodeling}. The linearized AC-OPF (Section~\ref{sec:opf}) is incorporated into the scheduler (Section~\ref{sec:scheduler}) and \ac{RT} controller (Section~\ref{sec:RTcontrol}). 

\section{Battery Modeling}
\label{sec:batmodeling}
In this section, we extend the linear battery models developed in our previous work \cite{fortenbacherPSCC14}. The resulting models capture the main relevant characteristics of battery systems and allow for tractable controller formulations. Note that these models are used within our controllers, but not used to simulate realistic battery systems. Recognizing that these models do not capture all battery dynamics, we use the DUALFOIL electrochemical battery model \cite{Fuller} to represent real batteries in the \ac{RT} simulations presented in Section~\ref{sec:results}.
\subsection{Efficiency}
\begin{figure}[!t]
	\centering
	\def\svgwidth{\columnwidth}
	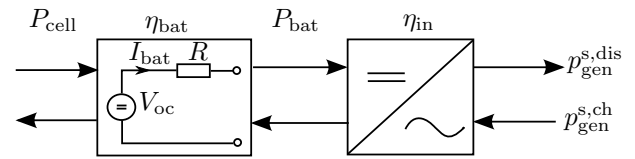
	\caption{Block diagram of a battery system consisting of a grid connected inverter (right block) and battery stack (left block).}
	\label{fig:effi}
\end{figure}

\begin{figure}[!t]
	\center
	\includegraphics[width=\columnwidth]{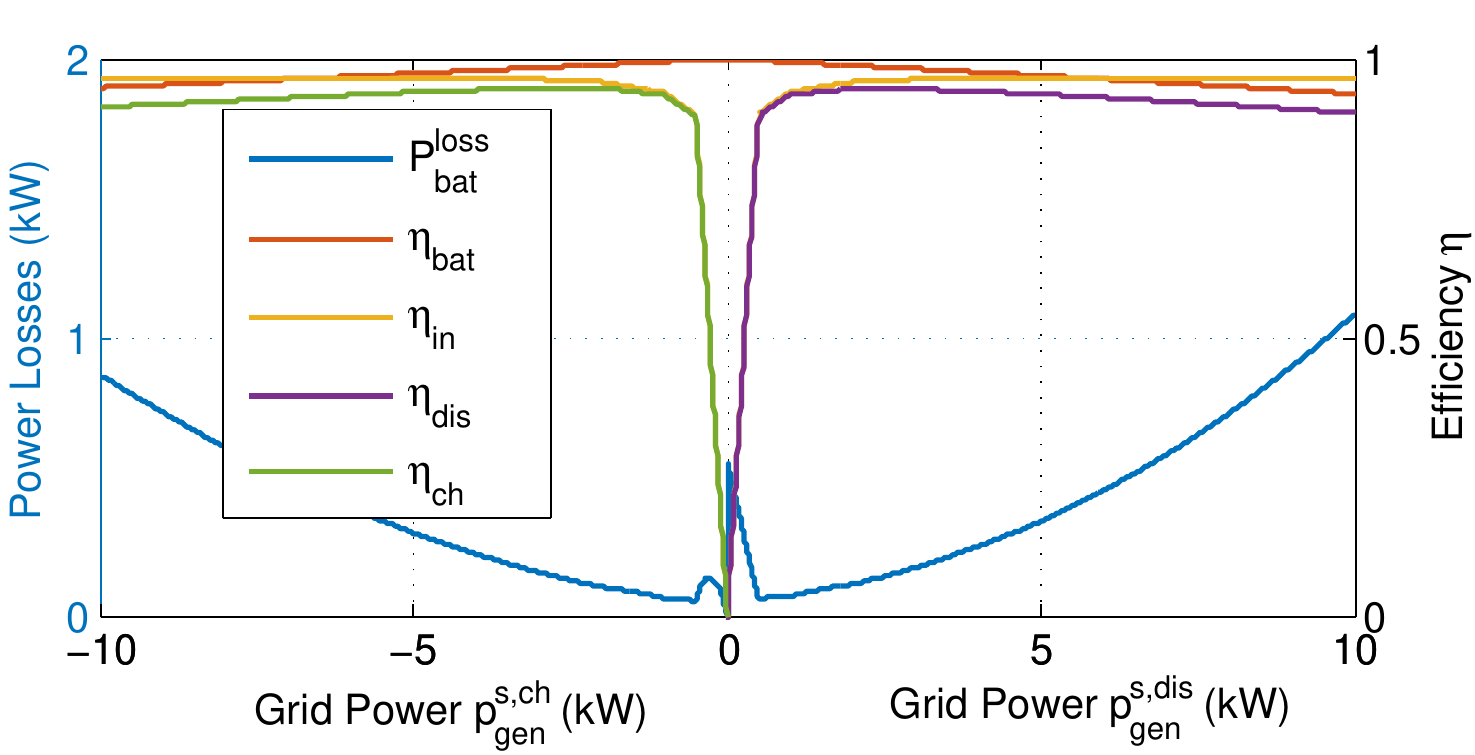}
	\caption{Battery system efficiencies and total power losses as a function of the grid power. The plot is calculated for a 10kW battery system with $R$= 10m$\Omega$,$V_{\mathrm{oc}}$= 42V, and a standard inverter efficiency curve from \cite{sma}.}
	\label{fig:powerlosses}
\end{figure}

In order to determine the cost optimal operation of multiple battery systems, we first need a model that describes the total efficiency of an individual battery system considering both the battery and inverter losses. The models developed in \cite{fortenbacherPSCC14} only included the battery losses so we extend them to include the inverter losses. Figure~\ref{fig:effi} shows a block diagram of a battery system. The crucial parameter that influences the battery efficiency \acs{etabat} is the battery's internal resistance $R$. Using a Thevenin circuit equivalent, the battery power $P_{\mathrm{bat}}$ is
\begin{equation} 
P_{\mathrm{bat}} = \acs{Vocsoc}\acs{ibat}-\acs{R}\acs{ibat}^2 \quad,
\label{eq:pbat}
\end{equation}
\noindent where \acs{Vocsoc} is the open circuit voltage (OCV). The battery efficiency for discharging currents $\acs{ibat} > 0$ is
\begin{equation}
\eta^{\mathrm{dis}}_{\mathrm{bat}}  =  \frac{P_\mathrm{bat}}{P_\mathrm{cell}}  = \frac{\acs{Vocsoc}\acs{ibat}-\acs{R}\acs{ibat}^2}{\acs{Vocsoc}\acs{ibat}} = 1 -  \frac{\acs{R}\acs{ibat}}{\acs{Vocsoc}}\label{eq:discharging} \quad,					
\end{equation}
\noindent where the internal cell power ${P_\mathrm{cell}}$ is referred to as the power input. The battery efficiency for charging currents $\acs{ibat} < 0$ is
\begin{equation}
\eta^{\mathrm{ch}}_{\mathrm{bat}} = \frac{P_\mathrm{cell}}{P_\mathrm{bat}} = 1 + \frac{\acs{R}\acs{ibat}}{\acs{Vocsoc}-\acs{R}\acs{ibat}} \quad.
\label{eq:charging}
\end{equation}
For $\acs{Vocsoc}>>\acs{R}\acs{ibat}$, we can approximate $\eta^{\mathrm{ch}}_{\mathrm{bat}} $ as follows:
\begin{equation}
\eta^{\mathrm{ch}}_{\mathrm{bat}} \approx \eta^{\mathrm{dis}}_{\mathrm{bat}} = 1 - \left|\frac{\acs{R}\acs{ibat}}{\acs{Vocsoc}}\right| \quad.
\label{eq:approx}
\end{equation}
The battery efficiency can be expressed as a function of \acs{pbat} by solving \eqref{eq:pbat} for \acs{ibat} and substituting the resulting expression into \eqref{eq:approx}
\begin{equation}
\acs{etabat} = 1 - \left|\frac{\acs{Vocsoc}-\sqrt{\acs{Vocsoc}^2-4\acs{R}\acs{pbat}}}{2\acs{Vocsoc}} \right| \quad.
\label{eq:losses}
\end{equation}
Including the inverter efficiency \acs{etain} the total losses are 
\begin{equation}
\acs{ploss}  = \begin{cases}
\left(\acs{etabat}^{-1}\acs{etain}^{-1} - 1 \right) p_{\mathrm{gen}}^{\mathrm{s,dis}} = \left(\eta_{\mathrm{dis}}^{ \ -1} - 1 \right) p_{\mathrm{gen}}^{\mathrm{s,dis}}  \\
(\acs{etabat}\acs{etain}-1)  p_{\mathrm{gen}}^{\mathrm{s,ch}}  =  \left(\eta_{\mathrm{ch}}-1\right) p_{\mathrm{gen}}^{\mathrm{s,ch}}, \\
\end{cases} \label{pbatloss}
\end{equation}
\noindent where ${p}_\mathrm{gen}^\mathrm{s,dis} \geq {0}$ and ${p}_\mathrm{gen}^\mathrm{s,ch} < {0} $ represent the discharging and charging grid powers. The multiplication of $\acs{etabat}$ and $\acs{etain}$ leads to a non-convex loss function, which is shown in Fig.~\ref{fig:powerlosses} (blue curve). The inverter efficiency is obtained from \cite{sma}. 
\subsection{Dynamics}
 
We also model fast battery dynamics. The models in \cite{fortenbacherPSCC14} expressed the state variable in terms of normalized charge, and here we express it in terms of energy for convenience.  

\subsubsection{Linear Basic Model}
\begin{figure}[!t]
	\centering
	\def\svgwidth{\columnwidth}
	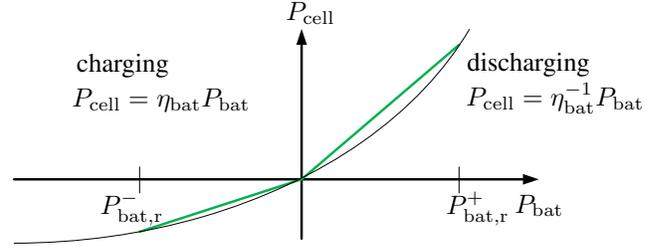
	\caption{Linear approximation of the cell power ${P_\mathrm{cell}}$ with respect to the battery power ${P_\mathrm{bat}}$.}
	\label{fig:linearloss}
\end{figure}
The first  model is an integrator model 
\begin{equation}
\dot{E}  =  -P_\mathrm{cell} \quad,
\end{equation}
\noindent where $E$ denotes the time-varying \ac{SoE}. As shown in Fig.~\ref{fig:linearloss}, $P_\mathrm{cell}$ is a nonlinear function of the battery power. To derive a linear model, we linearize $P_\mathrm{cell}$ between 0 and rated powers $P_{\mathrm{bat,r}}^-$, $P_{\mathrm{bat,r}}^+$ by evaluating \eqref{eq:losses} at those points. By including an averaged inverter efficiency $\bar{\eta}_{\mathrm{in}}$, we obtain
\begin{equation}
\dot{E} \approx \underbrace{[\underbrace{-\acs{etabat}(P_{\mathrm{bat,r}}^{+})^{-1}{\bar{\eta}_{\mathrm{in}}}^{\ -1}}_{-\eta_{\mathrm{dis}}^{ \ -1}}  \ \underbrace{- \acs{etabat}(P_{\mathrm{bat,r}}^{-})\bar{\eta}_\mathrm{in}}_{-\eta_{\mathrm{ch}}} ]}_{\vec{b}_\mathrm{bat}^T} \left[\begin{array}{c} p_{\mathrm{gen}}^{\mathrm{s,dis}} \\  p_{\mathrm{gen}}^{\mathrm{s,ch}}\end{array} \right] . \label{eq:linearbasic}
\end{equation}

\subsubsection{Linear Extended Model}
The linear extended model allows us to capture the rate capacity effect \cite{doyle} or charge relaxation effect \cite{Fuller1994}. This effect accounts for the ion diffusion in the electrolyte and reduces the accessible battery capacity for high battery powers. As we have presented in \cite{fortenbacherPSCC14} we capture this effect by modifying the KiBaM model \cite{Manwell1993399}. The model is
\begin{eqnarray}
\dot{\vec{x}}_\mathrm{E} &=& \underbrace{\left[
	\begin{array}[h]{rr}
	-\frac{\acs{cr}}{\acs{cw}} & \frac{\acs{cr}}{1-\acs{cw}} \\
	\frac{\acs{cr}}{\acs{cw}} & -\frac{\acs{cr}}{1-\acs{cw}} \\
	\end{array}\right]}_{\vec{A}}\vec{x}_\mathrm{E} +  \left[
\begin{array}[h]{c}
\vec{b}_\mathrm{bat}^T \\
\vec{0} \\
\end{array}\right]
\left[\begin{array}{c} p_{\mathrm{gen}}^{\mathrm{s,dis}} \\  p_{\mathrm{gen}}^{\mathrm{s,ch}}\end{array} \right] \nonumber   \\ 
E & = & x_{\mathrm{E}1}  + x_{\mathrm{E}2} \label{eq:extend}\quad,
\end{eqnarray}
\noindent where $\vec{x}_\mathrm{E}=[x_{\mathrm{E}1},x_{\mathrm{E}2}]^T $ denotes the state vector for two virtual wells that are interconnected. Energy can only be withdrawn from the available well $x_{\mathrm{E}1}$. The parameter \acs{cw} determines the  size of the wells and \acs{cr} is the inverse charge recovery time. 

\subsubsection{Comparison of models}
\begin{figure}[!t]
	\centering
	\includegraphics[width=\columnwidth]{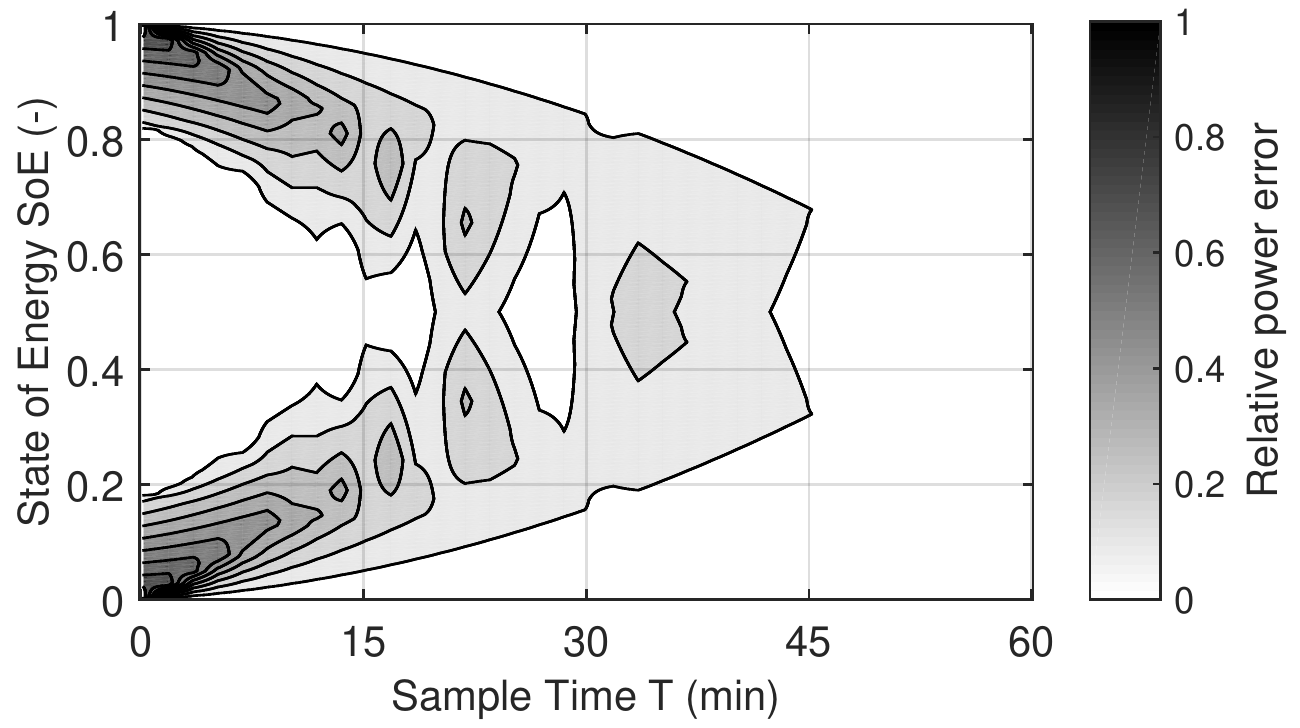}
	\caption{Relative battery power error of the linear basic model as compared to the linear extended model as a function of the sample time.}
	\label{fig:modelingerror}
\end{figure}

Figure~\ref{fig:modelingerror} demonstrates the merit of the linear extended model as compared to the linear basic model, which does not capture the rate capacity effect. The figure shows the relative error between the maximum power that could be applied to the system as modeled with the linear basic model as compared the linear extended model for different states of energy and as a function of the sample time used within discrete versions of the models. The linear basic model overestimates the maximum power available at high/low \ac{SoE} regimes for short sample times (seconds to minutes); however, it is as accurate as the linear extended model for sample times greater than or equal to 45 min. This means that for \ac{RT} control applications, the linear extended model is significantly more accurate than the linear basic model.

\subsection{Degradation}
\label{sec:convexhull}
\begin{figure}[!t]
	\center
	\includegraphics[width=\columnwidth]{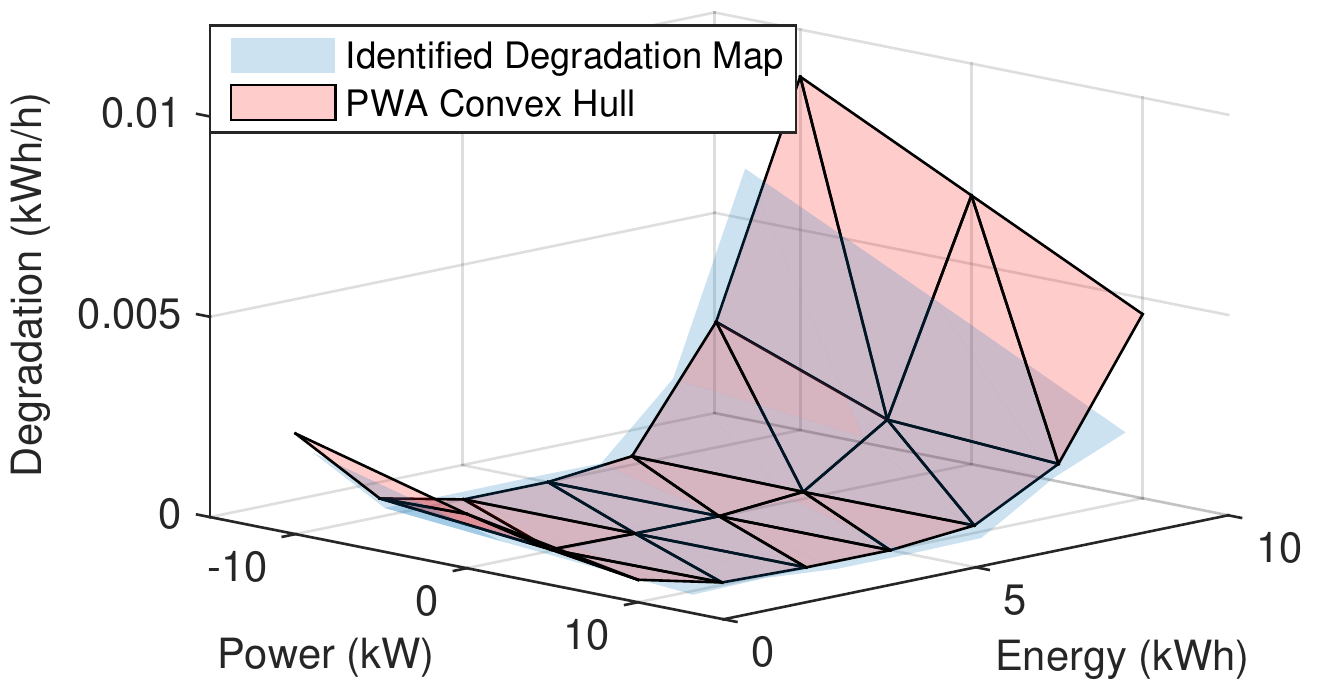}
	\caption{Illustration of the degradation map for a 10 kWh battery system ($c_{\mathrm{E}} = 10$kWh). The red surface is the \ac{PWA} convex hull \eqref{eq:hull} of the identified map (blue surface).}
	\label{fig:deg10kWh}
\end{figure}

Charge capacity loss in lithium ion (Li-ion) batteries is a slow process and hard to model. Among the contributors to capacity loss are two chemical side reactions that irreversibly transform cyclable ions into solids during battery operation. In \cite{fortenbacherPSCC14}, we presented a method to identify a stationary degradation process on an arbitrary battery usage pattern. The method produces a degradation map, where degradation is a function of the battery power and \ac{SoE}. Calendar life degradation is also included in the degradation map. Specifically, when the battery power is zero (i.e., the battery is resting), the degradation map in Fig.~\ref{fig:deg10kWh} shows non-zero degradation, which is a function of the \ac{SoE}. 

Unfortunately, degradation maps are in general nonconvex \cite{Forman2012}, such that we cannot apply efficient convex solvers in our optimization framework. As an extension to the work in \cite{fortenbacherPSCC14}, we compute the convex hull of the degradation map from \cite{fortenbacherPSCC14} using Delaunay triangulation \cite{delaunay}. Figure~\ref{fig:deg10kWh} shows the convex hull (red surface) of the identified degradation map (blue surface), assuming a LiCoO$_2$ graphite cell modeled with the DUALFOIL electrochemical battery model \cite{Fuller} configured as in Table~\ref{tab:dual}. The operating conditions used to determine the map are detailed in \cite{fortenbacherPSCC14}.  The rate constants \acs{rsc} and \acs{rsa} are tuned in such a way that after 4000 full cycles 80\% of the initial battery capacity remains. The map in \cite{fortenbacherPSCC14} is normalized with respect to the energy capacity $c_\mathrm{E}$ allowing us to scale the map for any battery size. By evaluating the plane parameters $\vec{a}_1,\vec{a}_2,\vec{a}_3 \in \mathbb{R}^{n_\mathrm{p}\times 1}$ of $n_\mathrm{p}$ triangles from the convex hull, we can define the following \ac{PWA} mapping for the normalized degradation rate $z/c_\mathrm{E}$:
\begin{equation}
\frac{z}{c_\mathrm{E}} = \max\left(\vec{a}_1 \ \frac{P_\mathrm{bat}}{c_\mathrm{E}} + \vec{a}_2 \ \frac{E}{c_\mathrm{E}} + \vec{a}_3\right)  \quad.
\label{eq:hullnorm}
\end{equation}
By multiplying \eqref{eq:hullnorm} by $c_\mathrm{E}$, we can express the degradation $z$ in absolute terms for any battery with energy capacity $c_\mathrm{E}$  as
\begin{equation}
z =  \max \left(\left[\vec{a}_1 \ \vec{a}_2 \ \vec{a}_3 \right]  \left[\begin{array}[h]{c}
	P_{\mathrm{bat}} \\
	E \\
	c_\mathrm{E}
\end{array}\right]  \right) \quad.
\label{eq:hull}
\end{equation}

\begin{table}[!t]
	\caption{DUALFOIL Configuration}
	\label{tab:dual}
	\centering
	\begin{tabular}{lll}
		\hline
		ambient conditions & isothermal  \\
		cathode & LiCoO$_2$ \\
		anode   & Graphite \\
		electrolyte & LiPF$_6$ \\
		\acs{Acell} & 1 m$^2$& \acl{Acell} corresponds to 30 Ah  \\
		            &         & \hspace{.2cm} for the cut-off voltage window \\ 
		\acs{rsc} & 1e-11 & \acl{rsc} \\
		\acs{rsa} & 1e-11 & \acl{rsa} \\
		$V_\mathrm{cut}^\mathrm{off}$ & 3.0 -- 4.2 V & cut-off voltage window\\
		\hline
	\end{tabular}
\end{table}

\section{Optimal Power Flow for Low Voltage Grids}
\label{sec:opf}
In our previous work \cite{fortenbacherPowerTech2015,fortenbacherPSCC16}, we derived a linearized  \ac{OPF} formulation. Based on the \ac{FBS} power flow method from \cite{Teng}, we linearly approximated voltage, power losses, and branch flow limits as linear functions of the nodal reactive and active power injections. The following assumptions were made. 
\begin{itemize}
	\item {We assume three-phase balanced radial LV networks. It is common to assume three-phase balanced LV networks in Europe \cite{cigre} since houses and PV inverters are connected to three phases; however, this assumption is less valid for North America. However, many studies make this assumption e.g.,~\cite{dall2016photovoltaic,DallAnese2014,ReyesChamorro2015,olivier2016}. The power flow equations we use could also be extended to handle unbalanced networks \cite{Teng} and can be incorporated in the proposed control framework. For example, \cite{dall2016photovoltaic} describes how one might extend a related power flow method.} 
	\item {We assume that \ac{LV} networks have high resistance over reactance ratios ($R/X\geq 2$) \cite{cigre}.}
	\item {We assume that voltage angle differences in \ac{LV} networks are small ($\leq 10^{\circ}$).}
\end{itemize}
We summarize the formulation here; for full details see \cite{fortenbacherPSCC16}.

\begin{figure}[!t]
	\centering
	\def\svgwidth{\columnwidth}
	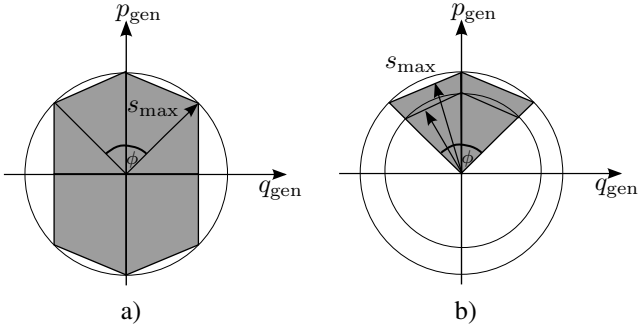
	\caption{Approximated reactive power capability areas a) circular-bounded b) $\cos\phi$-bounded. The polygonal convex regions can be described with the constraints (\ref{eq:FBOPF}h-l). \cite{fortenbacherPowerTech2015}}
	\label{fig:reactive}
\end{figure}

First, we give the generic formulation, before we specialize it to our scenario in Section~\ref{sec:control}.  We define following decision vector $\vec{x} = \left[\vec{p}_\mathrm{gen},\vec{q}_\mathrm{gen},\vec{p}_\mathrm{l}^\mathrm{p},\vec{p}_\mathrm{l}^\mathrm{q},\vec{v} \right]^T$, where $\vec{p}_\mathrm{gen} \in \mathbb{R}^{n_\mathrm{g} \times 1}$ and $\vec{q}_\mathrm{gen} \in \mathbb{R}^{n_\mathrm{g}\times 1 }$ are the active and reactive generator powers, where $n_g$ is the number of generators; $\vec{p}_\mathrm{l}^\mathrm{p}\in\mathbb{R}^{n_\mathrm{l}\times 1}$ and $\vec{p}_\mathrm{l}^\mathrm{q}\in\mathbb{R}^{n_\mathrm{l}\times 1}$ are the approximate active power losses due to active and reactive power injections (defined in \cite{fortenbacherPSCC16}), where $n_l$ is the number of lines; and $\vec{v}  \in \mathbb{R}^{n \times 1}$ is the voltage vector, where $n$ is the number of buses. Assuming positive linear costs $\vec{c}_\mathrm{p}$ for the active generator powers, the \ac{OPF} problem can be approximated as the following \ac{LP} problem:
\begin{equation}
\begin{array}{lllll}
\multicolumn{5}{l}{J^*=\displaystyle\min_{\vec{x}} \vec{c}_\mathrm{p}^T \vec{p}_\mathrm{gen} } \\
\hspace{0.0cm}\text{s.t.}    
& \text{(a)} & \multicolumn{3}{l}{\vec{1}^T\vec{C}_\mathrm{g}\vec{p}_\mathrm{gen} - \vec{1}^T\vec{p}_\mathrm{l}^\mathrm{p}-\vec{1}^T\vec{p}_\mathrm{l}^\mathrm{q}  = \vec{1}^T\vec{p}_\mathrm{d}  }\\
& \text{(b)} &  \multicolumn{3}{l}{\vec{B}_\mathrm{v} \left[\begin{array}{l} \vec{C}_\mathrm{g}\vec{p}_\mathrm{gen} \\ \vec{C}_\mathrm{g}\vec{q}_\mathrm{gen}  \end{array} \right] -\vec{v}  = \vec{B}_\mathrm{v}\left[\begin{array}{l} \vec{p}_\mathrm{d} \\ \vec{q}_\mathrm{d}  \end{array} \right] - \vec{v}_\mathrm{s} } \\
& \text{(c)} & \multicolumn{3}{l}{\vec{B}^1_{\mathrm{l}}\vec{p}_\mathrm{l}^\mathrm{p} - \vec{B}^2_{\mathrm{l}}\vec{C}_\mathrm{g}\vec{p}_\mathrm{gen} \geq \vec{B}^2_{\mathrm{l}} \vec{p}_\mathrm{d} + \vec{b}_{\mathrm{l}} } \\
& \text{(d)} & \multicolumn{3}{l}{\vec{B}^1_{\mathrm{l}}\vec{p}_\mathrm{l}^\mathrm{q} - \vec{B}^2_{\mathrm{l}}\vec{C}_\mathrm{g}\vec{q}_\mathrm{gen} \geq \vec{B}^2_{\mathrm{l}} \vec{q}_\mathrm{d} + \vec{b}_{\mathrm{l}} }  \\
& \text{(e)} & \multicolumn{3}{l}{-\vec{i}_\mathrm{b}^\mathrm{max} + \vec{B}_\mathrm{r}\vec{p}_\mathrm{d}\leq \vec{B}_\mathrm{r}\vec{C}_\mathrm{g}\vec{p}_\mathrm{gen} \leq \vec{i}_\mathrm{b}^\mathrm{max} + \vec{B}_\mathrm{r}\vec{p}_\mathrm{d}}\\
& \text{(f)} &\vec{v}_\mathrm{min} \leq \vec{v} \leq \vec{v}_\mathrm{max} \\
& \text{(g)} &\vec{p}_\mathrm{min} \leq \vec{p}_\mathrm{gen} \leq \vec{p}_\mathrm{max} \\
& \text{(h)} & \vec{p}_\mathrm{gen} +  \vec{A}_q\vec{q}_\mathrm{gen} \leq \vec{s}_{\mathrm{max}} \\
& \text{(i)} & \vec{p}_\mathrm{gen} - \vec{A}_q\vec{q}_\mathrm{gen} \leq \vec{s}_{\mathrm{max}} \\
& \text{(j)} & \vec{p}_\mathrm{gen} +  \vec{A}_q\vec{q}_\mathrm{gen} \geq -\vec{s}_{\mathrm{max}} \\
& \text{(k)} & \vec{p}_\mathrm{gen} - \vec{A}_q\vec{q}_\mathrm{gen} \geq -\vec{s}_{\mathrm{max}} \\
& \text{(l)} & -\vec{B}_q\vec{s}_{\mathrm{max}} \leq \vec{q}_\mathrm{gen} \leq\vec{B}_q\vec{s}_{\mathrm{max}} \quad,
\end{array}
\label{eq:FBOPF}
\end{equation}
\noindent where $\vec{p}_\mathrm{d} \in \mathbb{R}^{n \times 1}$ and $\vec{q}_\mathrm{d} \in \mathbb{R}^{n \times 1}$ are the active and reactive net load; $\vec{C}_\mathrm{g} \in \mathbb{R}^{n \times n_\mathrm{g}}$ maps generators to buses; and $\vec{B}_\mathrm{v}$, $\vec{B}_\mathrm{r}$, $\vec{A}_\mathrm{q}$, and $\vec{B}_\mathrm{q}$ are matrix parameters defined in \cite{fortenbacherPSCC16} and \cite{fortenbacherPowerTech2015}. Constraint (\ref{eq:FBOPF}a) enforces power balance in the grid. The voltage approximation is included in (\ref{eq:FBOPF}b), where each element of $\vec{v}_\mathrm{s} \in \mathbb{R}^{n_\mathrm{l}\times 1}$ is the slack bus voltage. The constraints (\ref{eq:FBOPF}c,d) incorporate epigraph formulations that are piecewise linear inner approximations of the power losses. We specify the matrix and vector parameters $\vec{B}^1_{\mathrm{l}}$, $\vec{B}^2_{\mathrm{l}}$, and $\vec{b}_{\mathrm{l}}$ to obtain a more compact form of the inequalities defined in \cite{fortenbacherPSCC16}. Constraint (\ref{eq:FBOPF}e) includes branch flow limits, where $\vec{i}_\mathrm{b} \in \mathbb{R}^{n_\mathrm{l}\times 1}$ are the branch flow currents. Constraints (\ref{eq:FBOPF}f,g) specify the lower and upper bounds for the voltage ($\vec{v}_\mathrm{min}, \vec{v}_\mathrm{max}$) and active generator powers  ($\vec{p}_\mathrm{min}, \vec{p}_\mathrm{max}$). Constraints (\ref{eq:FBOPF}h-l) approximate the generators' apparent power limits, where $\vec{s}_\mathrm{max}$ is the generators' maximum apparent power; specifically, we define circular-bounded and  $\cos\phi$-bounded active and reactive power settings by approximating the circular area/segments with convex sets \cite{fortenbacherPowerTech2015} that describe the polygons depicted in Fig~\ref{fig:reactive}.   

\section{Optimal Battery Operation}
\label{sec:control}
\subsection{Control Structure}
\begin{figure}[!t]
	\centering
	\def\svgwidth{\columnwidth}
	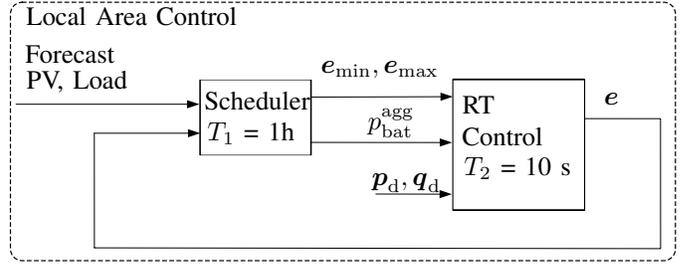
	\caption{Centralized control scheme consisting of a scheduler and an \ac{RT} controller. The scheduler uses robust \ac{MPC}. The \ac{RT} controller solves a \ac{RT}-\ac{OPF} using the storage allocation from the scheduler and \ac{RT} measurements.}
	\label{fig:scheduler}
\end{figure}

As shown in Fig.~\ref{fig:scheduler} our proposed control scheme consists of two control entities working at different time scales. We do not allow for PV curtailment and so the scheduler has to calculate future storage allocation bounds in terms of energy $\vec{e}_\mathrm{min},\vec{e}_\mathrm{max}$ and aggregated power $p_\mathrm{bat}^{\mathrm{agg}}$ that are feasible under the worst case \ac{PV} forecast scenarios. The \ac{RT} controller tries to control the storage units within these bounds, while minimizing network and storage losses and taking into account \ac{RT} measurements of the net load.          

\subsection{Scheduler}
\label{sec:scheduler}
We solve a robust multi-period \ac{OPF} subject to the worst case PV predictions. Since we solve the multi-period problems in a receding horizon fashion every hour with updated energy levels, the scheduler acts as a robust \ac{MPC}. 
\subsubsection{Incorporation of Distributed Storage}
Due to its high sample time $T_1=1$ hour, the scheduler cannot capture the detailed battery dynamics so we use the linear basic model~\eqref{eq:linearbasic}. The discrete version of \eqref{eq:linearbasic} for $n_s$ battery systems is    
\begin{equation}
\resizebox{1\hsize}{!}{$\vec{e}(k+1)  = \vec{I}^{n_s}\vec{e}(k) + 
\underbrace{T_1 \left[ \mathrm{diag}\{-\vec{\eta}_\mathrm{dis}^{-1}\} \ \mathrm{diag}\{-\vec{\eta}_\mathrm{ch}\} \right]}_{\vec{B}} \underbrace{\left[\begin{array}{l} \vec{p}_\mathrm{gen}^\mathrm{s,dis} \\ \vec{p}_\mathrm{gen}^\mathrm{s,ch} \end{array}\right]}_{\vec{p}_\mathrm{gen}^\mathrm{s}}$} ,						
\end{equation}
\noindent where $\vec{p}_\mathrm{gen}^\mathrm{s,dis} \geq \vec{0} \in \mathbb{R}^{n_s \times 1}$, $\vec{p}_\mathrm{gen}^\mathrm{s,ch} < \vec{0} \in \mathbb{R}^{n_s \times 1}$, 
$\vec{\eta}_\mathrm{dis},\vec{\eta}_\mathrm{ch} \in \mathbb{R}^{n_s \times 1}$ are the total discharging and charging efficiencies of the storage units, and $\vec{I}^{n_s}$ denotes the identity matrix of dimension $n_s$.
The complete SoE evolution $\vec{E} = [\vec{e}(1),...,\vec{e}(N-1)]^T$ for a time horizon of length $N$ is
\begin{equation}
\vec{E} =  \underbrace{\left[ \begin{array}{c} \vec{I}^{n_s} \\ \vdots \\ \vec{I}^{n_s}\end{array} \right]}_{\vec{S}_x}\vec{e}(0) + \underbrace{\left[\begin{array}{ccc}  \vec{B}  &  & \vec{0} \\ \vdots &\ddots \\ \vec{B} & \cdots & \vec{B} \end{array}\right]}_{\vec{S}_u}\underbrace{\left[\begin{array}{l} \vec{p}_\mathrm{gen}^\mathrm{s}(0) \\ \vdots \\  \vec{p}_\mathrm{gen}^\mathrm{s}(N-1) \end{array} \right]}_{\vec{U}} \quad, \label{eq:evo}
\end{equation}
\noindent where $\vec{e}(0)$ is the initial SoE vector.

\subsubsection{Incorporation of Battery Degradation}
In Section~\ref{sec:convexhull} we calculated the \ac{PWA} convex hull \eqref{eq:hull} of the degradation map. To incorporate degradation into the planning problem, we define helper decision variables $\vec{Z} = [\vec{z}(0),\hdots,\vec{z}(N-1) \in \mathbb{R}^{n_s\times 1}]^T$ and include the epigraphs of \eqref{eq:hull} as constraints in our problem. 
Specifically, we substitute $\vec{E}$ from \eqref{eq:evo} into an extended version of \eqref{eq:hull} that includes $n_s$ battery systems with energy capacities $\vec{c}_\mathrm{E} \in \mathbb{R}^{n_s \times 1}$ and obtain a degradation equation in terms of the control variable $\vec{U}$ as follows:
\begin{equation}
\left(\vec{A}_1+\vec{A}_2\vec{S}_u\right) \vec{U} + \vec{A}_z\vec{Z} \leq -\vec{A}_2\vec{S}_x\vec{e}_0-\vec{A}_3\vec{c}_\mathrm{E} \label{eq:batdeg}\quad,
\end{equation}
where $\vec{A}_1 = [\vec{I}^{N n_s}\otimes \vec{a}_1 ][\vec{I}^{N} \otimes [\vec{I}^{Nn_s} \vec{I}^{Nn_s}]]$, $\vec{A}_2 = [\vec{I}^{N n_s }\otimes \vec{a}_2]$, $\vec{A}_3 = \vec{I}^{N} \otimes [\vec{I}^{n_s} \otimes \vec{a}_3 ]$, and $\vec{A}_z = [\vec{I}^{N n_s}\otimes -\vec{1}^{n_{\mathrm{p}}\times 1} ]$. The operator $\otimes$ denotes the Kronecker product. 

\subsubsection{Robust Scheduling}
To solve a multi-period \ac{OPF} we first incorporate a temporally-coupled sequence of single period OPF problems \eqref{eq:FBOPF}. We extend the decision vector to $\vec{X} = [\vec{x}_0,\cdots,\vec{x}_{N-1}]^T$, where $\vec{p}_{\mathrm{gen},k} = [{p}_{\mathrm{net}}(k) \ \vec{p}_{\mathrm{gen}}^{\mathrm{s}}(k) ]^T \in \vec{x}_k$ and ${p}_{\mathrm{net}}(k)$ (i.e., the net power into or out of the LAC) is treated as a slack generator.
Since we do not allow for PV curtailment, we set 
\begin{equation}
\vec{p}_{\mathrm{d},k} = -\vec{C}_\mathrm{pv} (\hat{\vec{p}}_{\mathrm{gen},k}^{\mathrm{pv}}+\vec{w}_k3\vec{\sigma}_{\mathrm{pv},k}^T) + \hat{\vec{p}}_{\mathrm{ld},k}\quad,
\end{equation} 
\noindent where the PV and load forecasts are denoted by $\hat{\vec{p}}_{\mathrm{gen},k}^{\mathrm{pv}}$ and $\hat{\vec{p}}_{\mathrm{ld},k}$, and $\vec{C}_\mathrm{pv} \in \mathbb{R}^{n\times n_{\mathrm{pv}}}$ maps the $n_{\mathrm{pv}}$ \ac{PV} units to the buses. The standard deviation {$\vec{\sigma}_{\mathrm{pv},k}$ of the PV forecast error could be determined by using the results from \cite{Bacher}. We define $\vec{W} = [\vec{w}_0,\hdots,\vec{w}_{N-1}]^T$, which is a box-constrained uncertainty set where $\vec{-1}\leq \vec{W} \leq \vec{1}$. Note that we do not consider load forecast uncertainty, since absolute load forecast errors are generally much smaller than absolute PV forecast errors for a grid with high PV penetration. The scheduling problem is 
\begin{equation}
	\begin{array}{llll}
		\multicolumn{4}{l} {\displaystyle\min_{\vec{X},\vec{Z}} \ T_1 \left(\sum\limits_{k=0}^{N-1} \ {c}_{\mathrm{n}} {p}_{\mathrm{net}}(k)  + {c}_{\mathrm{d}} \vec{1}^T\vec{z}(k) \right)}\\
		 \text{s.t.} & \text{(a)} & \multicolumn{2}{l}{ -\vec{S}_x\vec{e}(0) \leq \vec{S}_u {\vec{U}} \leq \vec{c}_{\mathrm{E}} - \vec{S}_x\vec{e}(0) }\\
		  & & \multicolumn{2}{l}{\forall \vec{x}_k: \displaystyle\max_{-\vec{1} \leq \vec{W} \leq \vec{1}} \text{(\ref{eq:FBOPF}{a-c,e})} ,\text{(\ref{eq:FBOPF}{d,g-l}),{\eqref{eq:batdeg}}} \ ,}  
			\end{array}
	\label{eq:robustProb} 
\end{equation}
\noindent where $c_\mathrm{n}$ and $c_\mathrm{d}$ are cost parameters for the net power and the battery degradation. With this formulation, the PV infeed will be maximized since the optimizer minimizes $p_\mathrm{net}$. The constraint set (\ref{eq:robustProb}a) operates the storage devices at their minimum and maximum bounds.
To retrieve the robust counterpart of \eqref{eq:robustProb}, we eliminate $\vec{w}_k$ and use $|\vec{\sigma}_{\mathrm{pv},k}|$ within each constraint \cite{bental}, since \vec{X} does not multiply the random variable \cite{Lofberg2012}.
\subsubsection{Determination of Robust Storage Level Bounds}
To link the planning domain with the \ac{RT} domain, the scheduler sends robust allocation bounds for the \ac{RT} controller
\begin{eqnarray}
\vec{e}_\mathrm{min}& = &\min(\vec{e}(0),\vec{e}(1)) \\
\vec{e}_\mathrm{max}& = &\max(\vec{e}(0),\vec{e}(1)) \quad ,
\end{eqnarray}
\noindent where we evaluate $\vec{e}(1)$ with \eqref{eq:evo}. The scheduler also sends the aggregated battery setpoint $p_\mathrm{bat}^{\mathrm{agg}}$ to give the \ac{RT} controller the ability to operate the batteries at their individual optimal operation points
\begin{equation}
p_\mathrm{bat}^{\mathrm{agg}}  =  \vec{1}^T \vec{p}_\mathrm{gen}^{*\mathrm{s}}(1) \quad,
\end{equation}
where $\vec{p}_\mathrm{gen}^{*\mathrm{s}}(1)$ is a portion of the optimal decision vector of the robust problem \eqref{eq:robustProb}. 

\begin{figure}[!t]
	\centering
	\def\svgwidth{\columnwidth}
	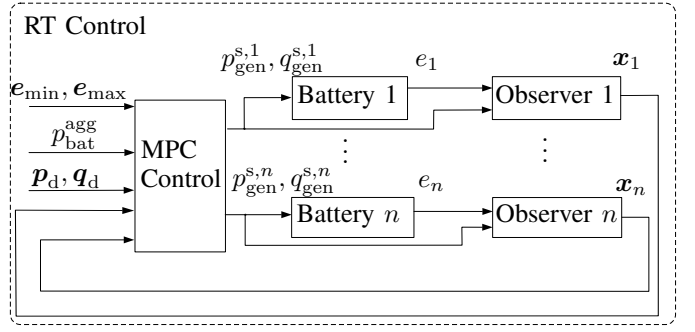
	\caption{Block diagram of the \ac{RT} controller.}
	\label{fig:rtControl}
\end{figure}  

\subsection{Real-Time Control}
\label{sec:RTcontrol}
As depicted in Fig.~\ref{fig:rtControl}, the \ac{RT} controller receives the storage allocation bounds and aggregated battery power setpoint. Using \ac{RT} measurements of the PV injections $\vec{p}_{\mathrm{gen}}^{\mathrm{pv}},\vec{q}_{\mathrm{gen}}^{\mathrm{pv}}$ and loads $\vec{p}_\mathrm{ld},\vec{q}_\mathrm{ld}$ to compute $\vec{p}_\mathrm{d},\vec{q}_\mathrm{d}$, it sets the reactive and active battery powers $\vec{p}_\mathrm{gen}^\mathrm{s},\vec{q}_\mathrm{gen}^\mathrm{s}$ by solving an RT-OPF. In particular, we solve an MPC problem for one step incorporating the linear extended battery model \eqref{eq:extend}, allowing us to estimate the power available from each individual battery system and utilize its full capacity potential. The decision vector is $\vec{u}_0 = [\vec{p}_\mathrm{gen}^\mathrm{s} \ \vec{q}_\mathrm{gen}^\mathrm{s} \ \vec{\lambda} \ \vec{p}_{\mathrm{bat}}^{\mathrm{loss}} \ \ y_\mathrm{set}^{\mathrm{agg}} \ \vec{x}_\mathrm{set}]^T$, where $\vec{\lambda}$ and $\vec{p}_{\mathrm{bat}}^{\mathrm{loss}}$ are helper variables to incorporate battery losses into the problem and $ y_\mathrm{set}^{\mathrm{agg}}$ and $\vec{x}_\mathrm{set}$ are helper variables to penalize deviations from the storage allocation bounds. The RT-OPF problem is
\begin{equation}
\begin{array}{lllll}
\displaystyle \min_{\vec{u}_0} 
&  \multicolumn{4}{l}{\underbrace{\vec{1}^T\vec{p}_{\mathrm{gen}}}_{\text{network losses}} + \underbrace{\vec{1}^T\vec{p}_{\mathrm{bat}}^{\mathrm{loss}} }_{\text{battery losses}} + \underbrace{ y_\mathrm{set}^{\mathrm{agg}} + \vec{1}^T \vec{x}_\mathrm{set}}_{\text{deviations from setpoint regions}}}\\
\hspace{0.2cm}\text{s.t.}   
& \multicolumn{4}{l}{(\ref{eq:FBOPF}\mathrm{a})-(\ref{eq:FBOPF}\mathrm{l})} \\
& \text{(a)} & \vec{X}_1& = & \vec{\Phi} \vec{X}_0 + \vec{H} \vec{p}_\mathrm{gen}^\mathrm{s} \\
& \text{(b)} & \vec{0} &\leq  &\vec{X}_1 \leq \vec{1}^{n_s \times 1} \otimes \left[ \begin{array}{c}
c_\mathrm{w} \\ 1-c_\mathrm{w} \end{array} \right] \\
& \text{(c)} & \multicolumn{3}{l}{-\operatorname{diag}\{\vec{m} \}\vec{C}\vec{X}_1 - \vec{I}^{n_s}\vec{x}_\mathrm{set} \leq  -\vec{e}_\mathrm{min}m}  \\
& \text{(d)} & \multicolumn{3}{l}{\operatorname{diag}\{\vec{m} \}\vec{C}\vec{X}_1 - \vec{I}^{n_s}\vec{x}_\mathrm{set} \leq  \vec{e}_\mathrm{max}m}  \\
& \text{(e)} & \multicolumn{3}{l}{ \vec{I}^{n_s}\vec{x}_\mathrm{set} \leq  \vec{0}} \\
& \text{(f)} & \multicolumn{3}{l}{ -m \vec{1}^T \vec{p}_{\mathrm{gen}}^{\mathrm{s}} - y_\mathrm{set}^{\mathrm{agg}}   \leq  -m {p}_{\mathrm{bat}}^{\mathrm{agg}}}   \\
& \text{(g)} & \multicolumn{3}{l}{  - y_\mathrm{set}^{\mathrm{agg}}   \leq  0 }  \\
& \text{(h)} & \multicolumn{3}{l}{\vec{p}_\mathrm{d} = -\vec{C}_\mathrm{pv} {\vec{p}}_{\mathrm{gen}}^{\mathrm{pv}} + {\vec{p}}_\mathrm{ld} } \\
& \text{(i)} & \multicolumn{3}{l}{\vec{q}_\mathrm{d} = -\vec{C}_\mathrm{pv} {\vec{q}}_{\mathrm{gen}}^{\mathrm{pv}} + {\vec{q}}_\mathrm{ld} } \\
& \text{(j)} & \multicolumn{3}{l}{-\vec{I}^{n_s} \vec{p}_\mathrm{gen}^\mathrm{s} + \left[\vec{I}^{n_s} \otimes \vec{p}_\mathrm{gen}^{\mathrm{s},P}\right]  \vec{\lambda} = \vec{0}  } \\
& \text{(k)} & \multicolumn{3}{l}{-\vec{I}^{n_s}  \vec{p}_{\mathrm{bat}}^{\mathrm{loss}} + \left[\vec{I}^{n_s} \otimes f(\vec{p}_\mathrm{gen}^{\mathrm{s},P})\right]  \vec{\lambda} = \vec{0}  } \\
& \text{(l)} & \multicolumn{3}{l}{\left[\vec{I}^{n_s} \otimes \vec{1}^{1\times n_s}\right]  \vec{\lambda} = \vec{1}  }\\
& \text{(m)} & \multicolumn{3}{l}{\vec{\lambda} = [\vec{\lambda}_1,\hdots,\vec{\lambda}_{n_s}]^T \ \forall \vec{\lambda}_i : \text{SOS2} \quad, }
\end{array}
\label{eq:mpc}
\end{equation}
\noindent where (\ref{eq:mpc}a) is the discrete version of \eqref{eq:extend} for multiple battery systems. Specifically, $\vec{X}_0$ is the initial battery system state and $\vec{X}_1$ is the state after one time step; \vec{\Phi} is the battery system dynamics matrix, $\vec{H}$ is the RT control input matrix, and $\vec{C}$ maps the internal states of each battery system to the SoE. Constraint (\ref{eq:mpc}b) avoids overspill of the battery capacity wells. Constraints (\ref{eq:mpc}c-e) define two regions and enable us to penalize deviations outside of the storage allocation bounds $\vec{e}_\mathrm{min},\vec{e}_\mathrm{max}$  with the penalty factor $m$, where $\operatorname{diag}\{\vec{m}\}$ is a square matrix with $m$ on the diagonal. We only penalize the aggregated power setpoint in the discharging direction with (\ref{eq:mpc}f,g), since overcharging can be dealt with in \ac{RT} as it will not result in line limit violations. This yields lower storage utilization and thus results in lower battery degradation.
Constraints (\ref{eq:mpc}h,i) incorporate the \ac{RT} measurements and (\ref{eq:mpc}j-m) represent nonconvex piecewise linear functions of the battery losses using an efficient \ac{SOS} formulation \cite{sos2}. In particular, we discretize \eqref{pbatloss} with $P$ supporting points $\vec{p}_\mathrm{gen}^{\mathrm{s},P}$ and evaluate \eqref{pbatloss} for each point and for each battery system. The sum of the elements in the set $\vec{\lambda}_i$ has to be equal to 1, which is enforced in (\ref{eq:mpc}l), and at most two elements in the set have to be consecutively non-zero (\ref{eq:mpc}m). This is referred to as an \ac{SOS}2 set. Constraint (\ref{eq:mpc}m) makes the problem a Mixed Integer Linear Programming (MILP) problem.  

We also implemented an observer to estimate the internal states $\vec{x}_\mathrm{E}$ of each individual battery system, since battery management systems typically provide only the SoE. Since the system \eqref{eq:extend} is detectable, \cite{observer} we can design following Luenberger observer
\begin{equation}
\dot{\hat{\vec{x}}}_\mathrm{E} = (\vec{A} -\vec{l} [1 \ 1])\hat{{\vec{x}}}_\mathrm{E} + \left[\begin{array}{cc}
\begin{array}[h]{c}
1  \\
0 \\
\end{array} & \vec{l}\end{array}\right] 
\left[
\begin{array}[h]{c}
P_\mathrm{bat}  \\
E \\
\end{array}\right] \quad,
\end{equation}
\noindent where $\vec{l} \in \mathbb{R}^{2\times 1}$ is chosen such that $(\vec{A}-\vec{l}[1 \ 1])$ is stable.


\section{Simulation Results}
\label{sec:results}
Through case studies we demonstrate the performance of our centralized control scheme. The parameters are listed in Table~\ref{tab:sim_parameters}. Figure~\ref{fig:scenario} shows the units within the test grid. We assess each control stage with simulations on different timescales. To study the RT controller, we simulate its performance on a sunny day and use the DUALFOIL electrochemical battery model~\cite{Doyle01011993} in place of a real battery system. We assume perfectly balanced cells, such that we can scale the model to emulate any required power and energy capacity. To study the impact of the scheduler on battery lifetimes, we need to consider a long time horizon. However, to reduce simulation times, we omit the \ac{RT} stage and run our scheduler with the simplified battery model and model the capacity loss with the convexified degradation maps. To assess battery lifetimes, we compute the capacity loss from the original degradation maps. 
\begin{figure}[!t]
	\centering
	\includegraphics[width=\columnwidth]{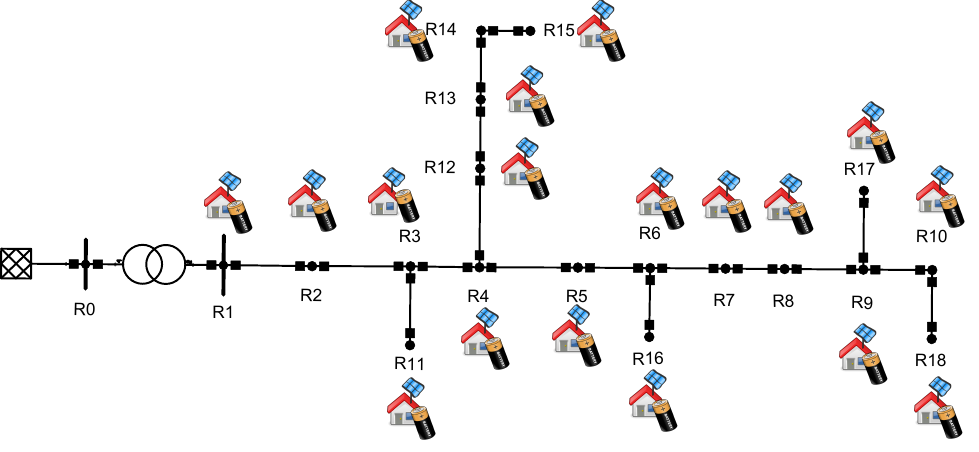}
	\caption{Modified Cigre test grid \cite{cigre} with PV generators (20kW) and battery storage units (10kVA/20kWh).}
	\label{fig:scenario}
\end{figure}
\begin{table}[!t]
	\centering
	\caption{Simulation parameters.}
	\begin{tabular}{lc} \hline
		Storage units & 18 \\
		Storage parameters &  $p_{\mathrm{gen}}^{\mathrm{s,max}}=$10 kW, $q_{\mathrm{gen}}^{\mathrm{s,max}}=$ 10 kVar, \\
						  & $c_\mathrm{E} =$20 kWh, $\eta_\mathrm{dis}$ = 0.91, $\eta_\mathrm{ch} $= 0.91  \\
						  & $V_\mathrm{oc}$ = 42 V, $R$ = 10 m$\Omega$ \\
		Inverter  & efficiency curve from \cite{sma} \\
		Deg.~model LiCoO2  & convexified degradation map from \cite{fortenbacherPSCC14} \\
		Deg.~model LiFePO4 & convexified degradation map from \cite{Forman2012} \\
		Scheduler horizon $N$ & 24 @ 1h sample time \\
		Net energy cost $c_\mathrm{n}$ &  100 \euro/MWh\\
		Degradation cost $c_\mathrm{d}$ & 400 \euro/kWh \\
		Supporting points &$P = 10$\\
		Penalization factor & $m = 1000$ \\
		PV units & 18 \\
		PV power & $p_{\mathrm{gen}}^{\mathrm{pv,max}}$= 20 kW ($\lambda$ = 1) \\
		Simulation horizon & 1 year\\
		Households & 18 @ 4kW (randomized profile) \\
		Grid & European \ac{LV} network \cite{cigre} \\
		Voltage Limits & $v_\mathrm{max} = 1.1$,$v_\mathrm{min} = 0.9$ \\
		\hline
	\end{tabular}
	\label{tab:sim_parameters}
\end{table}
\subsection{RT control performance}
In Fig.~\ref{fig:storage_dispatch} we show the \ac{RT} evolution of the battery energy levels. The gray boxes are the allocation bounds $\vec{e}_\mathrm{min},\vec{e}_\mathrm{max}$ from the scheduler. The scheduler uses the linear basic model, which does not include the rate capacity effect. However, the RT controller uses the linear extended model and so, in high state of energy regimes, it reduces the battery power to make the full capacity available, corresponding to a slower rise in energy levels. 
\begin{figure}[!t]
	\centering
	\includegraphics[width=\columnwidth]{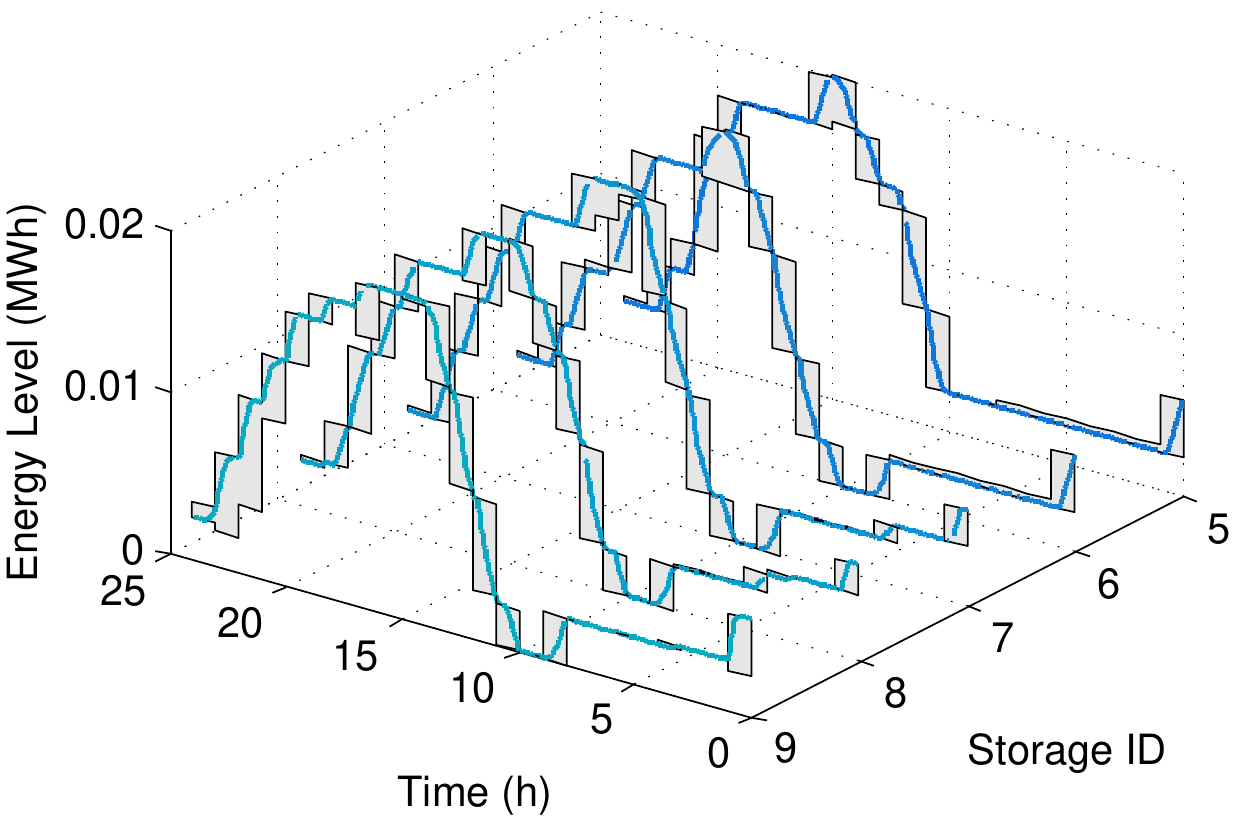}
	\caption{\ac{RT} evolution of the battery energy levels. The gray boxes are the storage allocation bounds from the scheduler, in which the RT controller can operate. For high state of energy regimes, the RT controller has to reduce the battery power to make the full capacity available.}
	\label{fig:storage_dispatch}
\end{figure}

To reduce battery system losses, the \ac{RT} controller dispatches the batteries to  loss-optimal operating points, while keeping them within the allocation bounds. This can be seen in Fig.~\ref{fig:rt_dispatch}, where the \ac{RT} controller sets the battery powers in a rolling fashion, resulting in switched battery operation.
\begin{figure}[!t]
	\centering
	\includegraphics[width=\columnwidth]{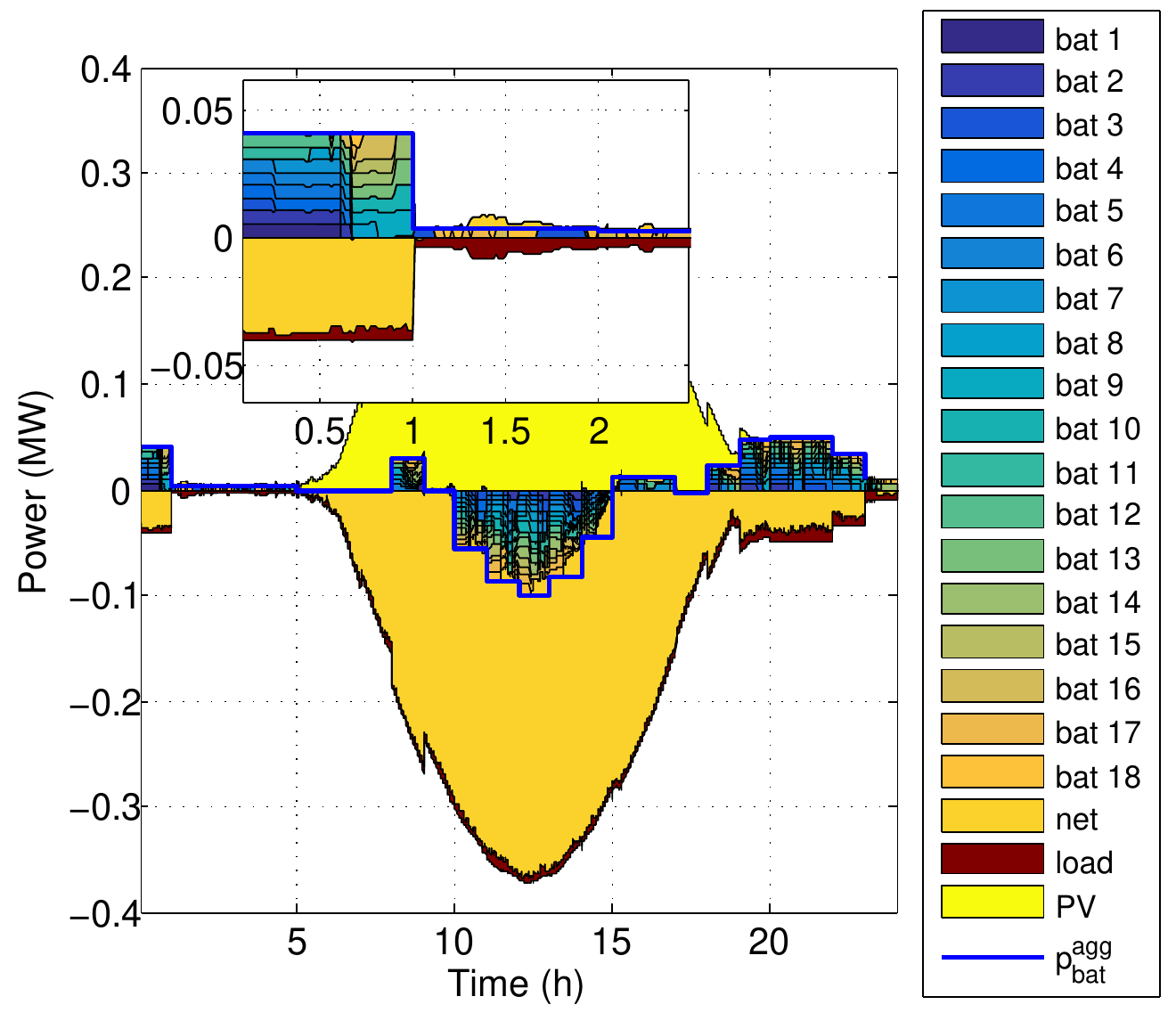}
	\caption{\ac{RT} dispatch for a one day simulation with a snapshot of the battery discharging phase. The blue curve shows the aggregated battery power signal from the scheduler. To reduce battery losses, the batteries are discharged (positive powers) and charged (negative powers) in a switched way.}
	\label{fig:rt_dispatch}
\end{figure}

Figure \ref{fig:vm_pf} shows compliance with the voltage and line flow constraints. 
Since we use a linear approximation of the power flow in our controllers, we calculate the exact power flows by running the full nonlinear AC power flow with all power injection variables set to the setpoints calculated by the \ac{RT} controller. As already shown in \cite{fortenbacherPSCC16}, the linear approximation leads to a solution that is always feasible. This we can see from Fig. \ref{fig:vm_pf}, since the maximum line capacity is not fully utilized and the allowable voltage band ($0.9-1.1$ pu) is not fully used. The mean absolute errors (MAEs) in voltage magnitudes/angles between the approximate power flow equations and the AC power flow equations are $1.06\times 10^{-3}$~pu and $5.94 \times 10^{-3}$~$^\circ$.
\begin{figure}[!t]
	\centering
	\includegraphics[width=\columnwidth]{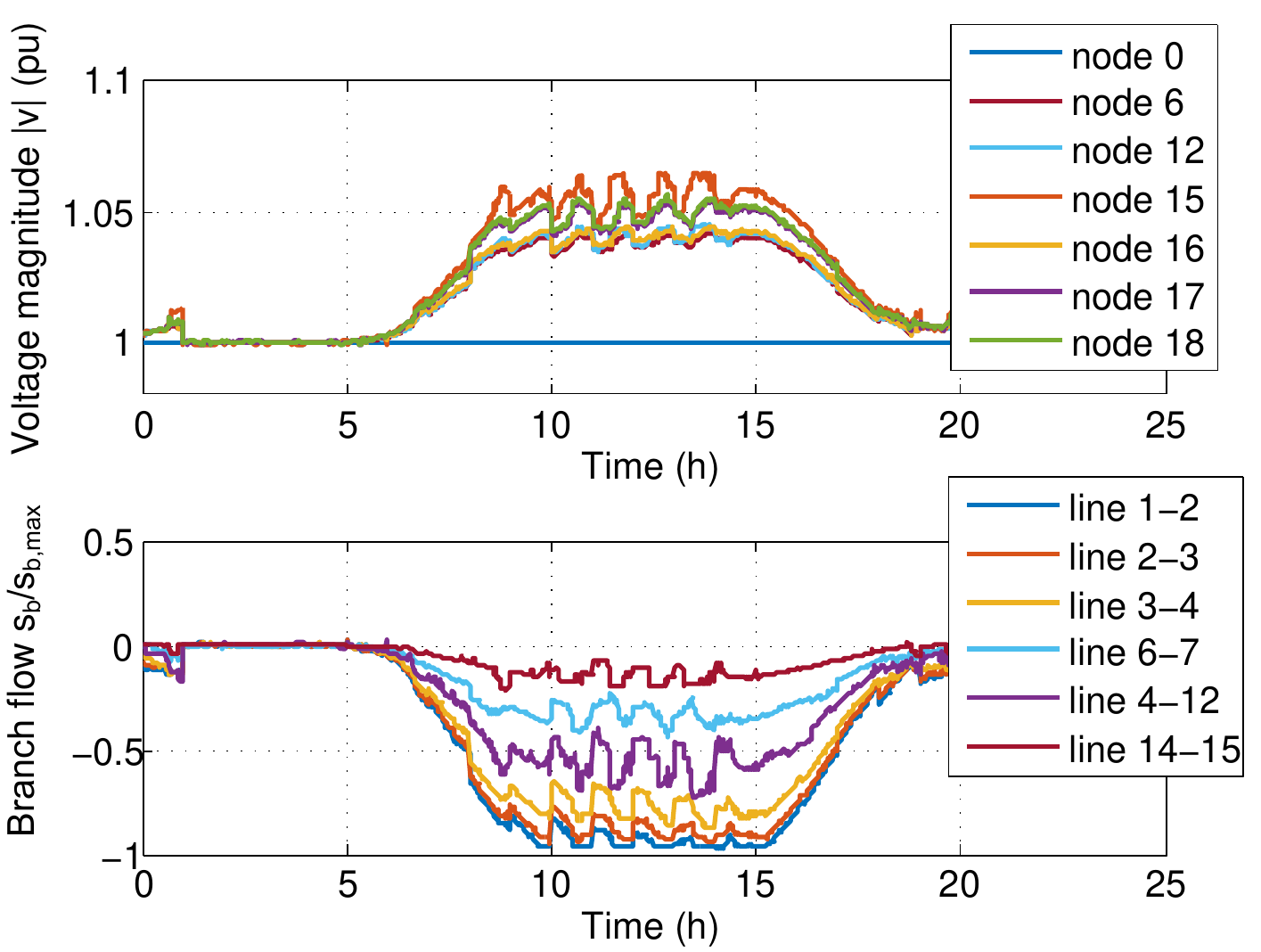}
	\caption{Voltage magnitudes and normalized branch power flows for a one day simulation with high PV infeed. The allowable voltage band is $\vec{v}_\mathrm{min}=0.9$ to $\vec{v}_\mathrm{max}=1.1$ pu.}
	\label{fig:vm_pf}
	\vspace{-0.5cm}
\end{figure}

To illustrate the merit of the detailed power loss model, we simulate the \ac{RT} controller using the full model in \eqref{eq:mpc} (referred to as the RT-MILP configuration) and using a simpler linear loss model, resulting in an \ac{LP} problem (referred to as the RT-LP configuration). As shown in Fig.~\ref{fig:rt_losses} the battery losses are 30\% smaller, if we use the RT-MILP configuration. Table~\ref{tab:results_RT} quantifies this finding. The loss reduction is mainly due to the reduced battery losses. Network losses are almost identical in both cases and are approximately two third of the total losses.
\begin{figure}[!t]
	\centering
	\includegraphics[width=\columnwidth]{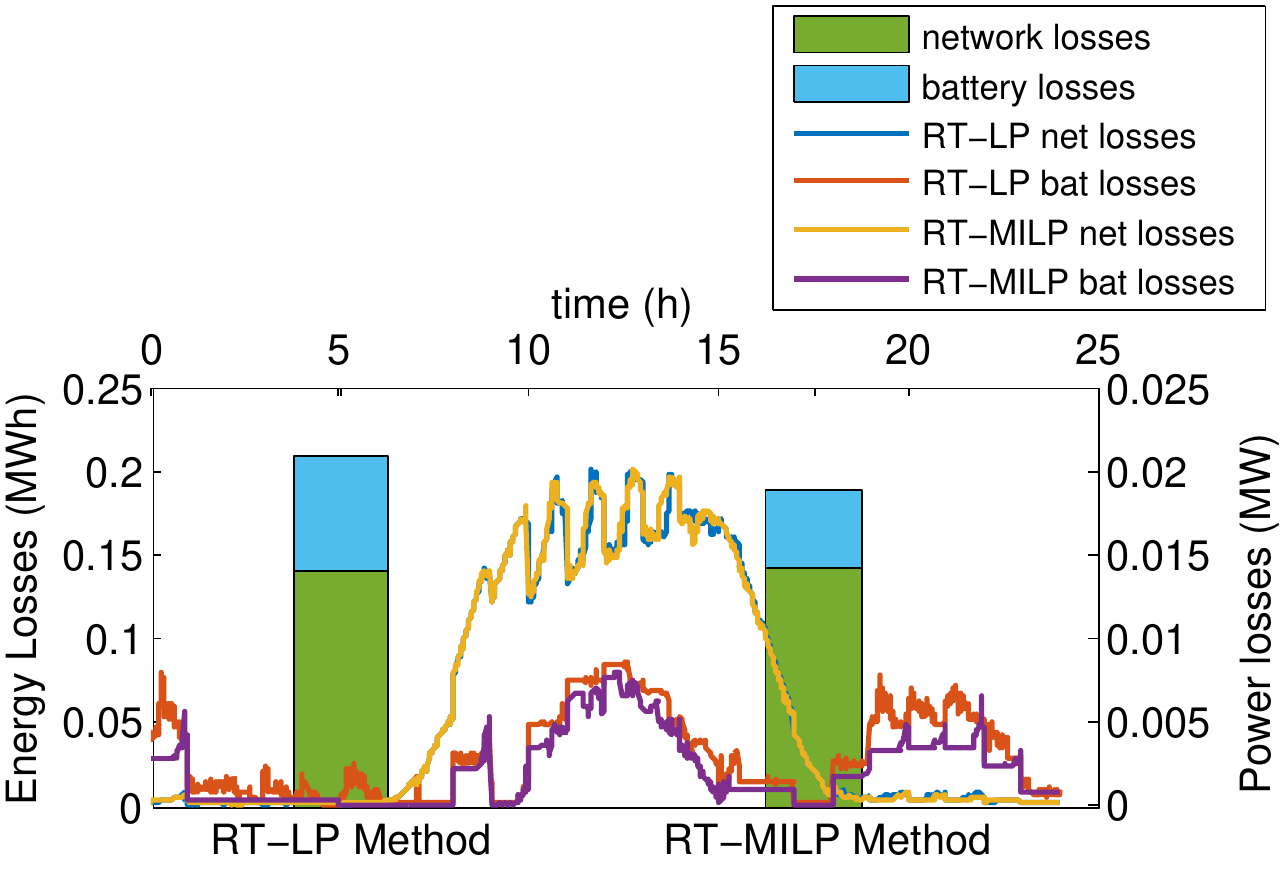}
	\caption{Battery and network losses for a sunny day. The bars show the energy losses as a function of the RT control configuration. The curves show power loss over the day for the different RT control configurations. With the detailed loss model (RT-MILP) the system incurs almost 30\% fewer losses.}
	\label{fig:rt_losses}
\end{figure}
Table~\ref{tab:results_simutimes} shows the computation time of each stage. We use the CPLEX \cite{cplex} optimization solver running on an Intel Core i7 computer. The computation time for the MILP problem is less than one second, which is reasonable for the \ac{RT} control stage. The MILP computation time for 200 devices is approximately 4 seconds and so the approach could be implemented in real-time.
\begin{table}[t]
	\centering
	\caption{Losses for different \ac{RT} control configurations}
	\begin{tabular}{lccc} \hline
		RT control   & battery  &network losses  & battery losses\\
		configuration & loss model & (kWh) & (kWh) \\
		\hline
		RT-MILP & nonconvex & 141.8 & 47.5 \\
		RT-LP & linear & 141.2 & 68.4  \\
		\hline
	\end{tabular}
	\label{tab:results_RT}
\end{table}
\begin{table}[t]
	\centering
	\caption{Problem sizes and computation times for the controllers}
%
%

\begin{tabular}{lcc}
\hline
& Scheduler & RT Controller \\
\hline
problem class & LP & MILP (18 SOSs) \\
number of constraints  & 14328 & 381 \\
number of decision variables  & 4392 & 327 \\
computation time $\pm 1\sigma$ (sec) & 2.85 $\pm$ 0.26 & 0.78 $\pm$ 0.12\\
\hline
\end{tabular}
	\label{tab:results_simutimes}
\end{table}
\subsection{Lifetime Assessment}
We next determine how the scheduler affects battery lifetimes, and compare two different battery technologies:  LiCoO2 and LiFePO4. For LiCoO2, we take the degradation map from \cite{fortenbacherPSCC14}. For LiFePO4, we discretize the analytic degradation function presented in \cite{Forman2012} to obtain a map. Both maps are convexified as described in Section~\ref{sec:convexhull} and are included in the scheduler. 
We also simulate the scheduler without using the degradation model, but constrain the minimum \ac{SoE} level to 5\% of the capacity.
In Fig.~\ref{fig:total_fade}, we compare the charge capacity loss over one year for the different battery technologies. The scheduler without the degradation model operates the batteries in low \ac{SoE} regimes. Those regimes result in high degradation, especially for the LiCoO2 system.  
\begin{figure}[!t]
	\centering
	\includegraphics[width=\columnwidth]{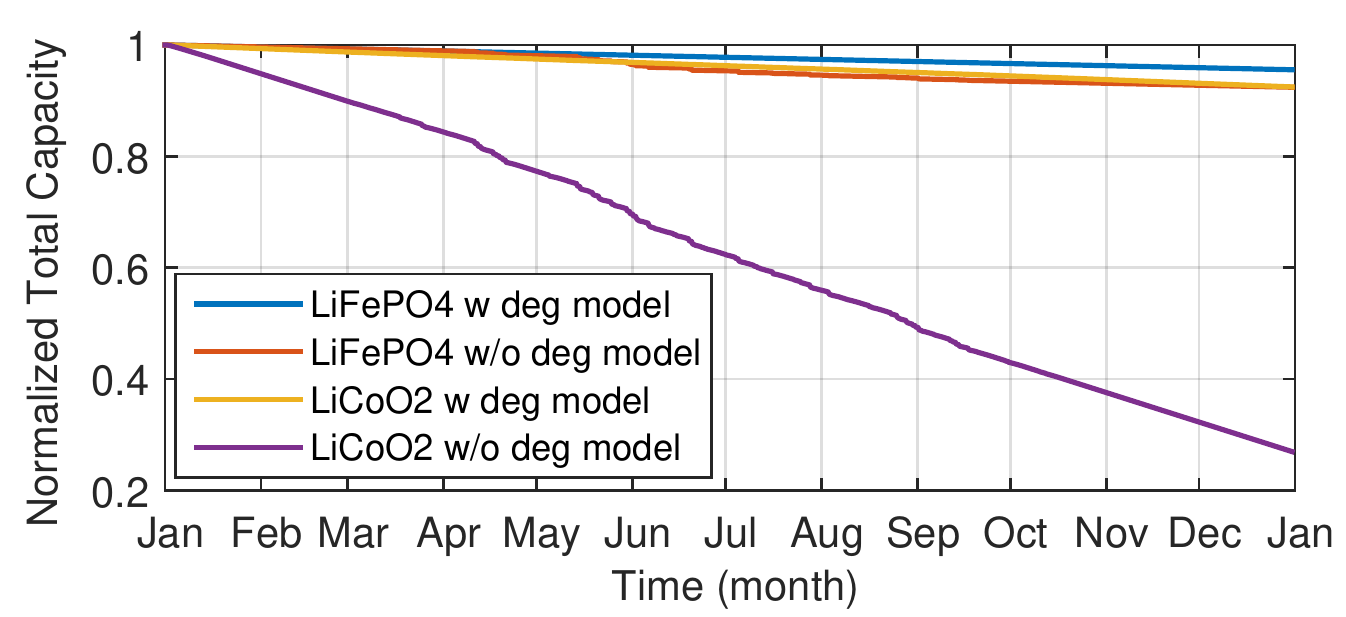}
	\caption{Normalized total battery capacity over one year for different battery technologies, with (w) and without (w/o) a degradation (deg) model. 
	}
	\label{fig:total_fade}
	\vspace{-0.3cm}
\end{figure}
In Fig.~\ref{fig:bat_fade}, we also show the charge capacity loss for each battery. Using degradation models, the scheduler balances degradation across the batteries, contributing to longer battery lifetimes (see Table~\ref{tab:results_scheduler}). We also extrapolated these results to obtain estimates for the number of full cycles and lifetimes assuming an end of life (EOL) criterion of 0.8, which means that 80\% of the initial capacity remains. We find that by using a degradation model, we can prolong the lifetime by a factor 10 for the LiCoO2 system and by a factor 2 for the LiFePO4.
\begin{figure}[!t]
	\centering
	\includegraphics[width=\columnwidth]{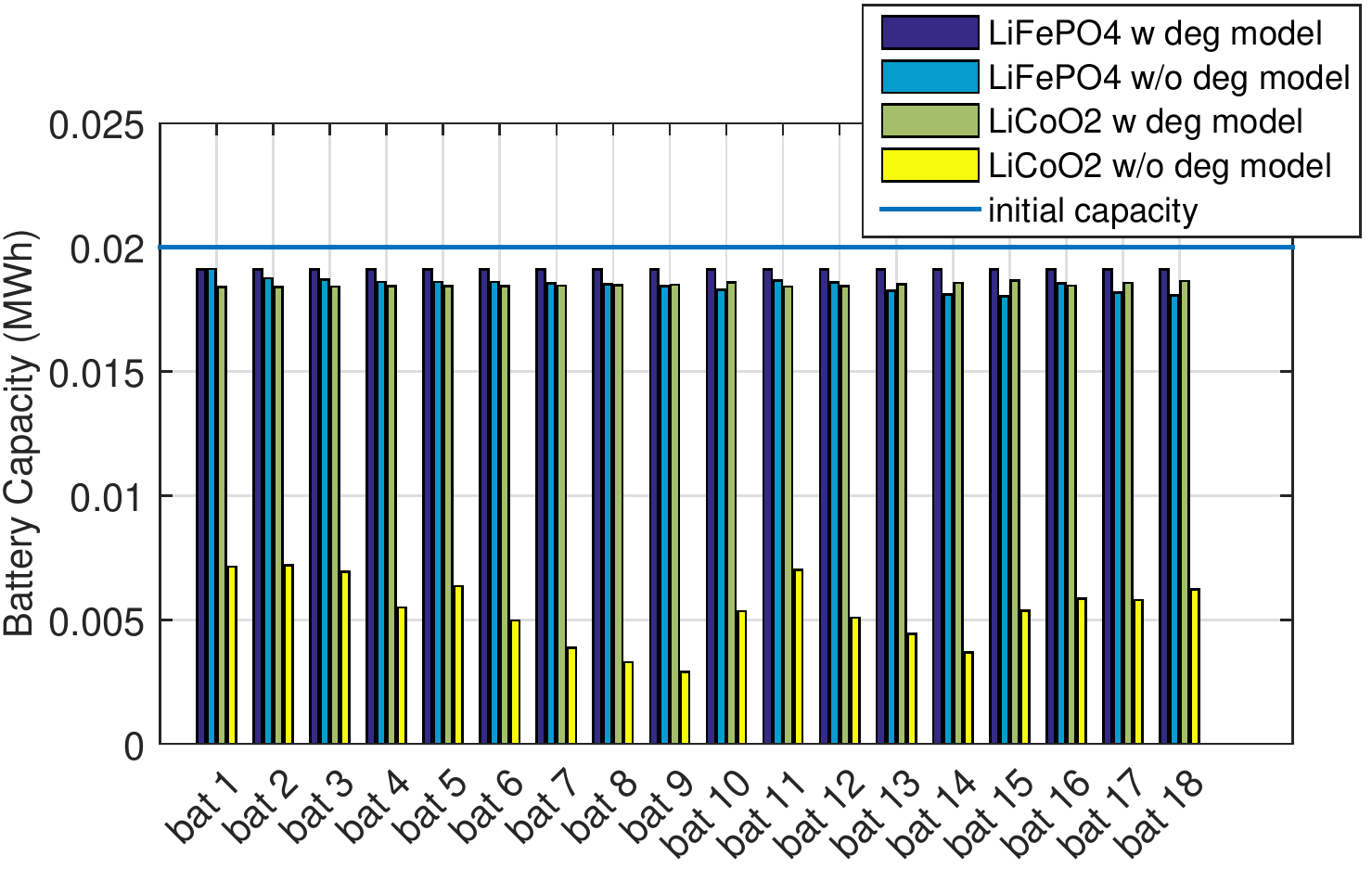}
	\caption{Remaining battery capacities for each battery (bat) after a one year simulation for different battery technologies, with (w) and without (w/o) a degradation (deg) model. The blue line shows the initial battery capacities.}
	\label{fig:bat_fade}
\end{figure}
\begin{table}[t]
	\centering
	\caption{Lifetime assessment for different battery technologies}
	\begin{tabular}{llcc} \hline
		Battery  & Configuration & Expected number & Expected lifetime  \\
		Technology & & of full cycles & (years) \\
		\hline
		LiFePO4					& w deg model &  2816 & 4.4 \\
		& w/o deg model & 1609 & 2.64 \\
		LiCoO2					& w deg model &   1652 & 2.63 \\
		& w/o deg model & 166 &  0.29 \\
		\hline
	\end{tabular}
	\label{tab:results_scheduler}
\end{table}

\section{conclusion}
\label{sec:conclusion}
In this paper, we present a novel two-stage centralized \ac{MPC} scheme for distributed battery storage to mitigate voltage and line flow violations that are induced by high PV penetrations in \ac{LV} grids. To link the  planning and real time domains, our control scheme consists of a robust scheduler and a \ac{RT} controller. This division enables planning using course PV and load forecasts on timescales of hours and RT control to manage dynamics and forecast error on timescales of seconds. Therefore, the scheduler uses simple battery models, while the RT controller uses detailed models. We incorporate a linearized \ac{AC-OPF} into both the scheduler and RT controller to reduce the computational complexity of the algorithms. To guarantee secure grid operation, the scheduler solves a robust multi-period \ac{OPF} taking the worst-case PV forecast into account. It produces robust feasible storage allocation bounds for the \ac{RT} controller, which maximizes PV utilization, while keeping battery degradation to a minimum by solving a single step OPF problem. We find that the control scheme can substantially reduce both battery losses and degradation. 

\bibliographystyle{IEEEtran}
\bibliography{literature}

\end{document}